\documentclass[letterpaper,twocolumn,10pt]{article}
\usepackage{usenix2025_SOUPS}

\usepackage{tikz}
\usepackage{amsmath}
\usepackage{multirow}

\usepackage{filecontents}

\begin{document}

\date{}

\title{\Large \bf AI For Privacy in Smart Homes: Exploring How Leveraging AI-Powered Smart Devices Enhances Privacy Protection}

\def\plainauthor{Author name(s) for PDF metadata. Don't forget to anonymize for submission!}

\author{
{\rm Wael Albayaydh}\\
wael.albayaydh@cs.ox.ac.uk\\
University of Oxford
\and
{\rm Ivan Flechais}\\
ivan.flechais@cs.ox.ac.uk\\
University of Oxford
 \and
\and
 {\rm Rui Zhao}\\
University of Oxford\\
rui.zhao@cs.ox.ac.uk
 \and
 {\rm Jood Albayaydh}\\
joud.bayaydh@institute-of-ai.org\\
Institute of AI
}

\maketitle

\begin{abstract}
Privacy concerns and fears of unauthorized access in smart home devices often stem from misunderstandings about how data is collected, used, and protected. This study explores how AI-powered tools can offer innovative privacy protections through clear, personalized, and contextual support to users. Through 23 in-depth interviews with users, AI developers, designers, and regulators, and using Grounded Theory analysis, we identified two key themes: \textit{Aspirations for AI-Enhanced Privacy} —how users perceive AI’s potential to empower them, address power imbalances, and improve ease of use— and \textit{AI Ethical, Security, and Regulatory Considerations}—challenges in strengthening data security, ensuring regulatory compliance, and promoting ethical AI practices.

Our findings contribute to the field by uncovering user aspirations for AI-driven privacy solutions, identifying key security and ethical challenges, and providing actionable recommendations for all stakeholders, particularly targeting smart device designers and AI developers, to guide the co-design of AI tools that enhance privacy protection in smart home devices. By bridging the gap between user expectations, AI capabilities, and regulatory frameworks, this work offers practical insights for shaping the future of privacy-conscious AI integration in smart homes.
\end{abstract}

\section{Introduction}\label{4th study intro}
The rise of smart home devices has improved convenience through automation and control but has also introduced significant privacy and security challenges, particularly in multi-user environments with varying levels of expertise. Our research explores \textbf{"How can AI-powered tools support privacy protection when integrated into smart home devices?"} By offering context-aware guidance, AI can empower users to safeguard their data while promoting transparency and usability in smart home privacy management.

AI-powered tools like ChatGPT~\footnote{\label{chatgpt}\href{https://chat.openai.com/}{\underline{\textcolor{blue}{https://chat.openai.com/}}}} assist both primary users (administrators) and passive users by simplifying privacy settings, detecting vulnerabilities, and providing actionable recommendations. Participants in our study emphasized the importance of AI in bridging knowledge gaps, ensuring usability, and strengthening security. We used Amazon Alexa as an example of a voice assistant, integrating it with ChatGPT to examine its potential for privacy protection. Using a Grounded Theory approach, we conducted semi-structured interviews with 23 participants, including administrators, passive users, AI developers, smart device designers, and data regulators.

Our findings highlight \textit{Aspirations for AI-Enhanced Privacy} and \textit{AI Ethical, Security, and Regulatory Considerations}. Participants emphasized the importance of AI tools offering real-time privacy alerts, intuitive data management, and personalized recommendations. AI was seen as especially valuable in educating users, particularly vulnerable groups like children, about privacy risks, with voice-command accessibility crucial for usability. Key security concerns, including encryption, anonymization, and strict access controls, were highlighted as essential for protecting all users. Participants also underscored the need for compliance with global privacy regulations, transparency in AI decision-making, and accountability for data breaches. Ethical considerations, such as mitigating algorithmic bias, ensuring informed consent, and providing tailored privacy controls for vulnerable groups, were also emphasized as critical for fostering trust.

Our contributions lie in three areas: (1) defining user expectations for AI-enhanced privacy management, (2) identifying security vulnerabilities and ethical risks, and (3) providing practical design recommendations for AI developers and smart device designers. By bridging the gap between user needs, AI capabilities, and regulatory frameworks, our research offers insights for developing privacy-conscious AI integration in smart homes.

This study highlights the potential of AI-powered privacy tools to enhance privacy, security, and usability while balancing automation with user control. Future research should focus on interdisciplinary collaborations to refine ethical AI guidelines and evaluate the long-term impact of AI privacy tools in real-world settings.

\section{Background} \label{background5th}

\subsection{Privacy Concerns with Smart Home}\label{Bystanders' Privacy Concerns with the Smart Home Devices}
Emerging threats such as data leaks, device misuse, and targeted burglaries pose significant privacy risks in smart homes, exacerbated by users' limited awareness of these vulnerabilities~\cite{zeng_end_2017, ali_cyber_2018}. Incidents like smart TVs recording conversations have been documented, raising alarm among users~\cite{wilson_smart_2015, malkin_what_2018, malkin_privacy_2019, wilson_benefits_2017, ahmed_understanding_2017, brand_survey_2020}. Such concerns have led some users to disconnect devices or limit their functionality as a precaution~\cite{ghiglieri_exploring_2017}. Despite these risks, many users are willing to compromise privacy in exchange for the perceived benefits of convenience and automation, often consenting to data sharing without fully understanding the implications~\cite{zheng_user_2018, albayaydh_exploring_2022, albayaydh_innovative_2023}. Research underscores the critical need for transparency and clarity in data collection practices, as they significantly influence user trust and engagement~\cite{naeini_privacy_2015, jakobi_catches_2017, tabassum_i_2019,windl_saferhome_2022}. While much of the literature focuses on active users, the privacy concerns of passive users—such as bystanders or cohabitants—remain underexplored. Studies have identified key factors affecting privacy in smart homes, including limited awareness, power imbalances, barriers to open privacy discussions, and gaps in regulation~\cite{albayaydh_examining_2023, albayaydh_exploring_2022, albayaydh_innovative_2023, al-alami_vulnerability_2017, bernd_bystanders_2019, bernd_balancing_2022}.

To address these challenges, recommendations have focused on adopting innovative technologies, such as AI-driven solutions, to enhance privacy protection for all users~\cite{marky_you_2020}. The evolving smart home landscape highlights the urgency of further research into privacy implications, particularly regarding asymmetric power dynamics within these contexts~\cite{madden_opinion_2019, mare_smart_2020, mcdonough_bystanders_2019, johnson_beyond_2020, marky_i_2020, ahmad_tangible_2020, albayaydh_exploring_2022, yao_privacy_2019, albayaydh_innovative_2023}. Developing effective strategies and technologies is essential to ensure equitable privacy protection for all user groups in the smart home ecosystem.

\subsection{Power Dynamics}\label{powerdynamics}
Smart home devices introduce notable privacy challenges, often amplified by disparities in users' knowledge, experience, and socio-economic status~\cite{bernd_balancing_2022, eckhoff_privacy_2018}. Socio-economic power imbalances exacerbate these tensions, limiting the ability of less privileged users to manage privacy~\cite{lau_alexa_2018,bernd_balancing_2022}. Passive users, such as cohabitants or employees, often defer to device owners, whose decisions may overlook others’ privacy due to existing power hierarchies~\cite{bernd_bystanders_2019, geeng_whos_2019}. Research highlights several contributing factors, including limited awareness, asymmetric power dynamics, contextual influences, and regulatory gaps~\cite{albayaydh_exploring_2022, albayaydh_innovative_2023, lee_information_2016, nissenbaum_privacy_2009, burke_contemporary_2006, benthall_contexts_2019, xue_who_2024}. These imbalances are particularly pronounced in hierarchical relationships, such as employer-employee interactions, where less powerful individuals face restricted rights and limited privacy agency~\cite{lupton_self-tracking_2014, gold_trending_2012, watkins_allen_workplace_2017, ball_workplace_2010, lee_electronic_2003}. In smart homes, device control often reflects underlying socio-economic structures, further intensifying privacy concerns~\cite{apthorpe_you_2022, j_kraemer_exploring_2019, geeng_whos_2019}.

Addressing these challenges is complex and requires innovative solutions. Efforts include technologies for detecting hidden cameras~\cite{liu_detecting_2018, senior_blindspot_2009}, signaling data collection activities~\cite{marky_you_2020, portnoff_somebodys_2015}, and fostering awareness through open discussions and user-centered interventions~\cite{albayaydh_exploring_2022, albayaydh_innovative_2023, zeng_understanding_2017}. These approaches aim to empower passive users, promote equitable privacy management, and mitigate the impact of power imbalances.

\subsection{Privacy Controls}\label{privacycontrol}
Prior research highlights the importance of contextualizing privacy preferences within users' daily lives, particularly in relation to social dynamics~\cite{forlizzi_service_2006, woodruff_would_2014}. Users are more likely to articulate clear privacy preferences when notifications are presented in an understandable and accessible manner~\cite{tabassum_i_2019, gerber_foxit_2018}. Strategies such as privacy labels that detail the types of data collected by devices have been proposed to aid decision-making~\cite{kelley_nutrition_2009}. However, much of the focus remains on implementing accountability measures and providing tools for informed decision-making rather than empowering users with direct control mechanisms~\cite{tabassum_i_2019, gerber_foxit_2018, emami-naeini_exploring_2019, kelley_nutrition_2009,felfernig_overview_2019}. Although service providers often aim to offer privacy controls, deceptive design techniques, commonly referred to as ``dark patterns'', frequently manipulate users into making unintended privacy choices~\cite{deceptive-patterns_deceptive_2020, mathur_dark_2019}. Some dark patterns even contravene data protection laws, and the EU-GDPR~\footnote{{GDPR: }{\href{https://gdpr-info.eu/}{\textcolor{blue}{\underline{\textcolor{blue}{https://gdpr-info.eu\label{gdpr}}}}}}} underscores the necessity of specific, informed, and unambiguous user consent~\cite{stern_improving_2014, vollmer_recital_2023}. Despite these regulatory efforts, usability challenges---such as limited awareness, poor findability, low comprehension, and a lack of trust---often lead users to accept default, permissive privacy settings~\cite{dhingra_default_2012, farke_are_2021, hsu_awareness_2020}.

To address these issues, researchers have explored innovative solutions such as user-friendly interfaces, privacy assistants, and AI tools designed to enhance usability and support informed decision-making~\cite{habib_evaluating_2022, acquisti_nudges_2018, lipford_understanding_2008, tsai_turtleguard_2017, pearman_user-friendly_2022}. These efforts aim to bridge the gap between user intentions and practical privacy management, enabling more effective and transparent control over personal data.

\subsection{AI Tools}\label{AI tools}
At the time of writing, there are several AI tools such as \textbf{ChatGPT\footref{chatgpt}} available in the market, each offering different capabilities. \textbf{Google Bard}\footnote{\href{https://bard.google.com}{\underline{\textcolor{blue}{https://bard.google.com}}}} is Google's conversational AI (ConvAI) that integrates with their wider ecosystem. \textbf{Microsoft's Copilot}\footnote{\href{https://copilot.microsoft.com/}{\underline{\textcolor{blue}{https://copilot.microsoft.com/}}}} enhances productivity tools with AI-driven features within the MS Office suite. \textbf{Claude by Anthropic}\footnote{\href{https://www.anthropic.com}{\underline{\textcolor{blue}{https://www.anthropic.com}}}} focuses on safe and user-friendly interactions. \textbf{Jasper AI}\footnote{\href{https://www.jasper.ai}{\underline{\textcolor{blue}{https://www.jasper.ai}}}} specializes in content generation for marketing. \textbf{OpenAI's Codex}\footnote{\href{https://openai.com/blog/openai-codex}{\underline{\textcolor{blue}{https://openai.com/blog/openai-codex}}}} assists programmers by generating and understanding code. \textbf{Hugging Face Transformers}\footnote{\href{https://huggingface.co/transformers}{\underline{\textcolor{blue}{https://huggingface.co/transformers}}}} provides a vast library of pre-trained models for various Natural language processing (NLP) tasks. \textbf{AI21 Labs' Jurassic-2}\footnote{\href{https://www.ai21.com}{\underline{\textcolor{blue}{https://www.ai21.com}}}} is designed for text generation and comprehension. \textbf{IBM Watson Assistant}\footnote{\href{https://www.ibm.com/cloud/watson-assistant}{\underline{\textcolor{blue}{https://www.ibm.com/cloud/watson-assistant}}}} is tailored for enterprise applications, automating customer service and business processes. \textbf{Replika}\footnote{\href{https://replika.ai}{\underline{\textcolor{blue}{https://replika.ai}}}} offers a more personal AI experience, focusing on companionship and mental health support. Finally, \textbf{Dialogflow by Google}\footnote{\href{https://cloud.google.com/dialogflow}{\underline{\textcolor{blue}{https://cloud.google.com/dialogflow}}}} powers conversational interfaces for websites, mobile apps, and IoT devices. These tools cater to various needs, from content creation and coding to customer service and personal companionship, leveraging AI to provide smart, interactive, and automated solutions across industries. In this study, we used ChatGPT as an example of AI integrated into smart home devices to enhance privacy protection.

\subsubsection{\textbf{AI and Privacy}}
Natural language processing (NLP), particularly Large Language Models (LLMs) like OpenAI’s ChatGPT, based on Generative Pre-trained Transformer (GPT) models, has gained significant prominence. These models are effective across various domains, including generating, explaining, and analyzing code~\cite{tian_is_2023}. Among them, GPT-4 has demonstrated exceptional performance in specialized fields such as medical and legal examinations and technical applications~\cite{bubeck_sparks_2023}. In the realms of cybersecurity and privacy, LLMs are being actively explored for identifying vulnerabilities in code~\cite{nong_chain--thought_2024, pearce_examining_2021} and applying domain-specific models for tasks like Named Entity Recognition and Multi-Class Classification~\cite{ranade_cybert_2021}. However, their advanced capabilities also pose risks, such as facilitating malicious activities by lowering barriers for hackers and automating harmful operations~\cite{fang_llm_2024, gupta_chatgpt_2023}.

The integration of ConvAI in smart home devices has enhanced functionality and usability~\footnote{\href{https://www.ibm.com/topics/conversational-ai}{\underline{\textcolor{blue}{https://www.ibm.com/topics/conversational-ai\label{conai}}}}}. 
Adding GenAI tools to these systems~\footnote{\href{https://www.whizzbridge.com/blog/20-best-generative-ai-tools}{\underline{\textcolor{blue}{https://www.whizzbridge.com/blog/20-best-generative-ai-tools\label{topai}}}}} introduces opportunities to address critical privacy concerns and navigate complex power dynamics within households. GenAI enables adaptive personalization, tailoring functionality and privacy settings to individual preferences. However, this personalization can align with pre-existing socio-economic and power structures, potentially reinforcing inequalities, and more research is needed to minimise this. Furthermore, GenAI has the potential for mitigating power imbalances and protecting privacy in smart homes, however this remains underexplored~\cite{cao_new_2023, azar_home_2023, tiwari_inbuilt_2023}.

This study addresses this gap by examining how tools like ChatGPT can support equitable privacy control, ensuring that AI-driven personalization enhances fairness. For example, GenAI could dynamically adjust permissions and adapt device behaviors as household roles or relationships evolve, such as when children gain more autonomy or when tenants’ privacy needs differ from the expectations of landlords. These functionalities, when carefully designed, offer a path to simultaneously protect privacy and maintain equitable power distribution among stakeholders. ConvAI~\cite{freed_conversational_2021} and GenAI~\cite{feuerriegel_generative_2024} share overlapping goals but serve distinct purposes in the AI landscape. ConvAI is designed to interact with users through dialogue, providing coherent and contextually relevant responses. Applications include chatbots~\footnote{\href{https://www.oracle.com/chatbots/what-is-a-chatbot/}{\underline{\textcolor{blue}{https://www.oracle.com/chatbots/what-is-a-chatbot\label{chatbot}}}}}, virtual assistants like Apple's Siri~\footnote{\href{https://www.apple.com/siri/}{\underline{\textcolor{blue}{https://www.apple.com/siri\label{siri}}}}} and Amazon's Alexa, and customer service interfaces. In contrast, GenAI focuses on creating original content, such as text, images, or music, using advanced models like transformers and GANs~\footnote{\href{https://www.geeksforgeeks.org/generative-adversarial-network-gan/}{\underline{\textcolor{blue}{https://www.geeksforgeeks.org/generative-adversarial-network-gan\label{gans}}}}}. In this study, we explore the intersection of these technologies within smart homes, focusing on how GenAI can support privacy protection and mediate power dynamics, providing users with equitable control over their smart contexts.

\subsubsection{\textbf{ChatGPT and Smart Home Devices}}
Integrating ChatGPT with smart home devices opens up a range of possibilities~\cite{cao_new_2023,azar_home_2023,tiwari_inbuilt_2023}, allowing users to control their environment using natural language processing and AI. Several smart home devices can be connected to ChatGPT, including smart speakers like Amazon Echo and Google Nest, smart thermostats like Nest Learning Thermostat~\footnote{\href{https://store.google.com/regionpicker}{\underline{\textcolor{blue}{https://store.google.com/regionpicker\label{gans}}}}}, smart lights like Philips Hue~\footnote{\href{https://www.philips-hue.com/en-us}{\underline{\textcolor{blue}{https://www.philips-hue.com/en-us}}}}, smart locks like August Smart Lock~\footnote{\href{https://www.cnet.com/home/security/best-smart-locks/}{\underline{\textcolor{blue}{https://www.cnet.com/home/security/best-smart-locks/}}}}, and smart security cameras like Arlo~\footnote{\href{https://www.arlo.com/en-us/}{\underline{\textcolor{blue}{https://www.arlo.com/en-us/}}}}. To achieve this integration, various protocols and technologies are necessary. These include APIs~\footnote{Application Programming Interfaces} provided by device manufacturers, which enable communication between ChatGPT and the devices. For example, Amazon Alexa Skills Kit~\footnote{\href{https://developer.amazon.com/en-US/alexa/alexa-skills-kit}{\underline{\textcolor{blue}{https://developer.amazon.com/en-US/alexa/alexa-skills-kit}}}}
 (ASK) or Google Assistant SDK~\footnote{\href{https://developers.google.com/assistant/sdk}{\underline{\textcolor{blue}{https://developers.google.com/assistant/sdk}}}} can be used to create custom skills or actions that allow ChatGPT to control smart home devices. 

At the time of writing, this integration is not yet widely available on the market in a plug-and-play manner. However, developers can create custom integrations using available APIs and SDKs to connect AI services with specific smart home devices. The level of integration and control depends on the capabilities provided by the device manufacturers and the creativity of the developers.

\subsection{\textbf{AI and Data Protection Regulations}}\label{ai regulation}
Data protection in smart home devices has been explored extensively in various studies~\cite{badii_smart_2020,torre_framework_2016,varadharajan_data_2016,perera_big_2015,chaudhuri_internet_2016,shahid_data_2022,bastos_gdpr_2018,ioannidou_general_2021}. Our review of regulations reveals a variety of legal frameworks worldwide (see Table {\textcolor{blue}{\ref{tab:privacy-regworld}}}). The GDPR, often referred to as the \textit{“Golden Standard”} in Data Protection Law~\cite{pierozzi_data_2018}, has influenced many countries, including France, where the French National Data Protection Commission (CNIL~\footnote{\href{https://www.cnil.fr/en/home}{\underline{\textcolor{blue}{https://www.cnil.fr/en/home}}}\label{cnil}}) fined Google for GDPR violations concerning transparency and consent~\footnote{\href{https://www.cnil.fr/en/gdpr-developers-guide}{\underline{\textcolor{blue}{https://www.cnil.fr/en/gdpr-developers-guide}}}\label{googlecnil}}. As nations increasingly adopt GDPR-inspired data protection laws, there has been a global shift toward recognizing data protection as a fundamental right, ensuring enforceable privacy rights, and establishing independent supervisory authorities~\footnote{\href{https://unctad.org}{\underline{\textcolor{blue}{https://unctad.org}}}}~\cite{bennett_european_2018,knijnenburg_modern_2022,skyflow_how_2023}.

However, many countries still face challenges in creating comprehensive data protection laws. For example, while Canada's PIPEDA~\footnote{\href{https://www.justice.gc.ca/eng/csj-sjc/pa-lprp/pa-lprp.html}{\underline{\textcolor{blue}{https://www.justice.gc.ca/eng/csj-sjc/pa-lprp/pa-lprp.html\label{canadaPIPEDA}}}}}, Turkey’s DPL~\footnote{\href{https://www.kvkk.gov.tr/Icerik/6649/Personal-Data-Protection-Law}{\underline{\textcolor{blue}{https://www.kvkk.gov.tr/Icerik/6649/Personal-Data-Protection-Law\label{turkeydpl}}}}}, and Brazil’s LGPD~\footnote{\href{https://www.dataguidance.com/jurisdiction/brazil}{\underline{\textcolor{blue}{https://www.dataguidance.com/jurisdiction/brazil\label{brazilLGPD}}}}} focus on personal data protection but lack specific provisions for AI in smart homes. Similarly, in the U.S., data protection laws vary by state, with no unified federal law, although specific laws exist for areas like health data (HIPAA~\footnote{\href{https://www.cdc.gov/phlp/publications/topic/hipaa.html}{\underline{\textcolor{blue}{https://www.cdc.gov/phlp/publications/topic/hipaa.html}}}}), financial data (GLBA~\footnote{\href{https://www.ftc.gov/business-guidance/privacy-security}{\underline{\textcolor{blue}{https://www.ftc.gov/business-guidance/privacy-security}}}}), and children's online privacy (COPPA~\footnote{\href{https://www.ftc.gov/business-guidance/privacy-security/childrens-privacy}{\underline{\textcolor{blue}{https://www.ftc.gov/business-guidance/privacy-security/childrens-privacy}}}}). In countries like China (PIPL~\footnote{\href{https://www.dlapiperdataprotection.com/index.html?c=CN&t=law}{\underline{\textcolor{blue}{https://www.dlapiperdataprotection.com/index.html?c=CN\&t=law\label{chinaPIPL}}}}}), India (DPDP~\footnote{\href{https://prsindia.org/billtrack/digital-personal-data-protection-bill-2023/}{\underline{\textcolor{blue}{https://prsindia.org/billtrack/digital-personal-data-protection-bill-2023\label{indialaw}}}}}), and Malaysia (PDPA~\footnote{\href{https://www.termsfeed.com/blog/malaysia-pdpa/}{\underline{\textcolor{blue}{https://www.termsfeed.com/blog/malaysia-pdpa\label{malasyialaw}}}}}), GDPR-inspired provisions such as consent and the right to be forgotten are present, but there are no explicit regulations addressing AI use in smart homes.

Artificial Intelligence (AI) has become a transformative force, shaping industries and daily life through its ability to process massive amounts of data and make predictions based on human behavior~\cite{phiri_exponential_2023}. While AI holds immense potential, it also raises concerns related to privacy, data protection, and ethical issues. AI's dependence on personal data, particularly in areas like personal assistants and facial recognition, poses privacy risks and challenges regarding data transparency and profiling~\cite{shaelou_challenges_2023}. Furthermore, the increasing application of AI in everyday technology highlights the need for legal frameworks that govern its use. In 2024, several U.S. states, including Colorado~\footnote{\href{https://leg.colorado.gov/bills/hb23-1288}{\underline{\textcolor{blue}{https://leg.colorado.gov/bills/hb23-1288}}}}, Florida~\footnote{\href{https://www.flsenate.gov/Session/Bill/2023/7026}{\underline{\textcolor{blue}{https://www.flsenate.gov/Session/Bill/2023/7026}}}}, and Maryland~\footnote{\href{https://mgaleg.maryland.gov/mgawebsite/Legislation/Details/hb0209}{\underline{\textcolor{blue}{https://mgaleg.maryland.gov/mgawebsite/Legislation/Details/hb0209}}}}, enacted laws addressing algorithmic discrimination, AI in education, and state systems, yet there are no specific regulations for AI in smart homes. Existing data protection laws like GDPR, CCPA~\footnote{\href{https://iapp.org/resources/topics/ccpa-and-cpra/}{\underline{\textcolor{blue}{https://iapp.org/resources/topics/ccpa-and-cpra\label{ccpa}}}}}, and cybersecurity laws such as California’s IoT Security Law apply to AI in smart homes indirectly, but future legislation is needed to better address AI's role in these technologies.


\section{Methodology}
To address the research question, and consistent with methods in previous qualitative studies on user behavior with emerging technologies~\cite{albayaydh_exploring_2022, albayaydh_co-designing_2024, bernd_balancing_2022}, we conducted a Grounded Theory qualitative study to explore how AI can support privacy protection in smart home devices. Grounded Theory \cite{strauss_grounded_1997,glaser_discovery_1967} was chosen as it is effective for researching areas with limited prior study, allowing for theory construction through data collection, coding, and inductive reasoning. This approach is particularly suitable for examining data protection interactions, and it facilitates comprehensive insights into behaviors and beliefs. In this study, we focused on data protection concerns in smart home devices and the potential role of AI tools in managing privacy equitably, addressing power dynamics.

We engaged participants from different roles in the smart home ecosystem, including AI developers, designers, regulators, and users (both passive users and administrators). A total of 23 participants were interviewed to gather their perspectives on AI for privacy protection in smart devices, with a focus on how power dynamics affect privacy management. The study also demonstrated how a smart device (Alexa), integrated with ChatGPT~\footnote{\href{https://www.analyticsinsight.net/chatgpt/how-to-integrate-chatgpt-to-alexa-a-guide}{\underline{\textcolor{blue}{https://www.analyticsinsight.net/chatgpt/how-to-integrate-chatgpt-to-alexa-a-guide}}}}, could enhance privacy protection. Participants interacted with the Alexa device to understand how AI tools can aid in equitable privacy management. For other devices without direct ChatGPT integration, we asked participants to imagine such integration and respond accordingly.

While the sample size was small, the diverse range of participants allowed for capturing various perspectives on power dynamics in privacy management. The interview questions explored the impact of power imbalances on privacy decisions. Although the brief interviews may limit insights into more complex dynamics, the diverse participant pool and device demonstrations provided valuable understanding of how AI tools could influence privacy and power relationships in smart homes. Our findings highlight the need for equitable privacy management and emphasize the importance of considering power dynamics when integrating AI in smart devices. Further research with larger and more diverse participant groups would deepen our understanding of these dynamics.

We use the term \textit{Passive Users} to refer to users of smart home devices without control over device functions or access to monitored data. These users rely on existing devices without administrative access. The term \textit{Admins} refers to participants who control and use smart home devices. \textit{AI Developers} refers to those with experience in AI development for smart home devices and privacy. \textit{Designers} refers to those designing smart home devices, and \textit{Regulators} represents policymakers responsible for data protection.

\subsection{Recruitment}
We developed our screening questionnaire survey with the following criteria~\footnote{\href{https://drive.google.com/file/d/1zMz5HVhkY73-prBSPfKlTcMspOMDrp57/view?usp=sharing}{\underline{\textcolor{blue}{Download a copy of the screening questionnaire survey\label{survey}}}}}: participants must be able to communicate in English, consent to participate in the interviews, and consent to being audio-recorded. Additionally, participants should have experience using ChatGPT. For AI developers, and designers, a minimum of 2 years of experience with smart home device design (e.g., Product Development, UX Design, UI Design, Software, Firmware, etc.) is required. Both passive users and admins should have a good understanding of smart home devices, and regulators should have at least 2 years of experience with data protection policies. In this study, we did not interview children due to the sensitivity of engaging with this vulnerable user group. Instead, we spoke with their adult family members (parents and siblings) to understand their views on children's privacy, security, and usability needs concerning smart home devices.

To recruit AI developers, designers, and regulators, we advertised the study~\footnote{\href{https://drive.google.com/file/d/1_aohQQUGEhkQViExLMUefOg_Y6s2NFzQ/view?usp=sharing}{\underline{\textcolor{blue}{Download a copy of the study poster\label{studyposter}}}}} on LinkedIn and social media expert groups across the UK, Europe, and the USA (e.g., IBM AI, IoT, GDPR groups). We also conducted online searches (e.g., Google, LinkedIn) to identify and invite potential candidates. However, recruiting these professionals proved difficult due to the sensitivity of data protection topics and Non-Disclosure Agreements in place in many organizations \cite{keane_gdpr_2018, calder_eu_2020}. To overcome this, we used snowball sampling \cite{goodman_snowball_1961}, initially recruiting two AI developers, one smart device designer, and two regulators. This led to additional participants being introduced over time. For user recruitment, we posted in specialized groups on Facebook and other platforms, connecting with four users directly and a further 12 through snowball sampling. Although personal ownership of smart home devices can provide unique insights, our inclusion criteria required participants to have substantial knowledge of smart devices without necessarily owning one. This approach ensured our participant sample was as broad as possible to facilitate different perspectives on the potential benefits and risks associated with AI integration. Participants' competence with smart devices was defined using the Dreyfus model of skill acquisition \cite{dreyfus_five-stage_1980}.

We contacted 34 potential candidates through LinkedIn, email, and phone, and 29 agreed to complete our screening questionnaire. Ultimately, 23 participants were recruited: six passive users, six admins, four AI developers, three designers, and four regulators, representing six organizations and 12 households. Some AI developers, designers, and regulators came from the same organizations, but no measures were taken to exclude them as they were interviewed individually and had no significant bias or conflicts of interest. The 12 users were from separate households to minimize potential bias. We use labels to refer to the participants, e.g., AI01 for the first AI developer, D01 for the first designer, and so on. Recruitment was iterative: after initially interviewing and analysing the data from 15 participants, we identified gaps and interviewed eight more, bringing the total to 23. Interviews explored privacy protection in AI-integrated smart devices, gathering insights from users, developers, designers, and regulators. Prior to each interview, participants interacted with an integrated Alexa with ChatGPT for up to 30 minutes to better understand the research concept. For detailed demographic information of the participants, refer to Table-{\textcolor{blue}{\ref{Table:Demographic Information of Participants}}}.


\subsection{Data Collection}\label{Focus-group and approach}
We conducted semi-structured interviews remotely via Zoom and Facebook Messenger. The interviews were audio-recorded and followed a prepared list of questions, with flexibility to add follow-ups or skip topics as needed. At the start of each interview, we demonstrated Alexa integrated with ChatGPT, allowing participants to interact for up to 30 minutes before the discussion. The interviews, conducted in English between June and August 2024 by a trained researcher, provided rich qualitative data on AI-driven privacy in smart homes. Participants volunteered without compensation. Prior to the full study, a pilot study with five semi-structured interviews ensured the clarity of the questions and helped identify potential issues. This pilot study, involving participants relevant to each group, also included feedback from two researchers, resulting in no significant changes to the interview scripts.

\subsubsection{\textbf{Interviews}}

We interviewed five key stakeholder groups: passive users, admins, AI developers, designers, and regulators, gathering insights on smart home devices, data collection, AI-driven privacy enhancements, and security concerns. We also addressed vulnerable users, such as children, to assess privacy and security safeguards. To minimize bias, interviews began with general experience-related questions before delving into privacy and security issues. Passive users discussed their smart devices, the data collected, and their understanding of data protection rights, along with their views on AI tools like ChatGPT for privacy control and concerns about AI’s role in smart homes, including power imbalances. Discussions also covered risks for vulnerable users, such as children~\footnote{\href{https://drive.google.com/file/d/13ojwaJhDz4q67qfke1ysGn2EWAlAP6TU/view?usp=sharing}{\underline{\textcolor{blue}{Download the passive users interview guide\label{studyposter}}}}}.

Admins explained how they manage smart home data for themselves and passive users, including their awareness of data protection rights and approaches to power dynamics. We also examined AI’s role in security and privacy, with special attention to vulnerable users like children and the elderly~\footnote{\href{https://drive.google.com/file/d/1hfjTCimV8UhxLLCBEG2q_M7dtFdGVfCd/view?usp=sharing}{\underline{\textcolor{blue}{Download the admin interview guide\label{studyposter}}}}}. AI developers shared their understanding of smart home data collection and AI-driven privacy protections, addressing challenges in integrating AI for security and how it could safeguard vulnerable users~\footnote{\href{https://drive.google.com/file/d/1-Z4pZFSMEUWxC5XEZeB7at6IF2tVJZ0i/view?usp=sharing}{\underline{\textcolor{blue}{Download the AI developers interview guide\label{studyposter}}}}}. Designers described the smart devices they create, the data collected, and AI’s role in privacy, focusing on AI integration and protections for passive and vulnerable users~\footnote{\href{https://drive.google.com/file/d/1QMrSEHHNI0igm6Y5uquqI-jYo1p_cNPX/view?usp=sharing}{\underline{\textcolor{blue}{Download the designers interview guide\label{studyposter}}}}}. Finally, regulators discussed smart home data protection and AI regulations, highlighting ChatGPT’s role in privacy and concerns regarding policies for protecting passive and vulnerable users~\footnote{\href{https://drive.google.com/file/d/1wv5SYHOo5vyWwrDmi0F-2U_Ky4hxMloo/view?usp=sharing}{\underline{\textcolor{blue}{Download the regulators interview guide\label{studyposter}}}}}.

\subsection{\textbf{Analysis}}
Following the Grounded Theory procedure by Strauss and Corbin \cite{strauss_grounded_1997,glaser_discovery_1967}, we transcribed and analyzed the 23 semi-structured interviews. Two researchers, Author 1 (primary researcher) and Author 2 (principal investigator), independently conducted the initial coding of all interview scripts. During this process, Author 2 sought clarifications, and Author 1 annotated the scripts for context. This initial coding resulted in 246 codes, forming the study codebook (see Table-\textcolor{blue}{\ref{Table: Categories and Themes}}). We applied these codes to the other interviews through constant comparison, adding new codes as necessary. The codes were then grouped into themes (axial coding) and categories (selective coding) based on their dimensions and properties. Axial coding allowed us to integrate perspectives from all user groups: passive users, admins, AI developers, designers, and regulators. The researchers held regular meetings to discuss emerging codes and categorize them. Data saturation~\cite{corbin_basics_2014,guest_how_2006,collingridge_quality_2008}, a key methodological principle in Grounded Theory indicating that no further data collection is needed, was observed separately for each group: between the 5th and 6th interviews for passive users and admins, the 3rd and 4th for AI developers, the 2nd and 3rd for designers, and the 3rd and 4th for regulators. After finalizing the codebook, we tested its reliability through Grounded Theory triangulation \cite{jonsen_using_2009}. We randomly selected 5 participants (one from each group) to review the codes and themes, confirming consensus and providing additional insights. These were incorporated into the codes, resulting in 258 codes organized into the themes in Section \textcolor{blue}{\ref{findings}}. To verify reliability, we tested for inter-rater reliability, finding an average Cohen's kappa coefficient ($\kappa$) of 0.86, indicating near-perfect agreement\cite{mchugh_interrater_2012}. The analyzed interview material totaled 4 hours and 6 minutes for passive users (41 minutes per interview), 4 hours and 36 minutes for admins (46 minutes per interview), 3 hours and 12 minutes for AI developers (48 minutes per interview), 2 hours and 12 minutes for designers (44 minutes per interview), and 2 hours and 44 minutes for regulators (41 minutes per interview).

\subsection{Limitations}
Like all qualitative research, this study has several limitations:

\textit{Language Barriers}: Interviews were conducted in English, which was not the first language for some participants. This occasionally led to imprecise language and difficulty in expressing ideas clearly. However, we believe the impact on data collection and analysis was minimal, as validation through triangulation did not reveal significant issues \cite{baxter_understanding_2015,dell_yours_2012}.

\textit{Researcher Skill and Bias}: The quality of qualitative research depends on the researchers’ skills and may be influenced by personal biases, with inexperienced researchers potentially missing important data \cite{koskei_role_2015}. To mitigate this, the primary researcher conducting all 23 interviews is a trained professional, focusing on neutral and open-ended questioning to minimize bias.

\textit{Self-Reporting Bias}: Self-reporting bias is a common challenge in qualitative research~\cite{jupp_sage_2006}, as participants may forget details or provide inaccurate responses. They may also alter their answers based on what they believe the researchers expect, introducing bias~\cite{trueman_structured_2015}. To enhance validity and mitigate this bias, proactive measures were implemented, including open-ended questions and probing to gather more detailed responses.

\textit{Recruitment Challenges}: Recruiting participants was challenging due to the sensitive nature of the research. We were unable to engage AI developers and designers from large organizations, which limited our sample. Ethical concerns and the sensitivity of involving children prevented direct interviews, so insights were gathered from adult family members. Future studies should include children while ensuring all ethical considerations are addressed. Participant responses may have been biased due to concerns about corporate reputation. To mitigate this, we assured participants of our stringent security and privacy measures, including data anonymization and encryption in compliance with GDPR.

\textit{Generalization and Scope of Findings}: This qualitative study focused on gaining in-depth insights rather than generalizing to a broader population. We reached saturation within each participant group, ensuring our analysis reflected participant views. Our findings, based on the perspectives of 23 participants, specifically explored ChatGPT as a GenAI tool, with the aim of probing this quickly evolving space. Future research, involving larger samples, will expand on generalizations and explore other AI tools.


\subsection{\textbf{Ethical considerations}}

Our study was approved by our institution's Research Ethics Committee [Approval: xxxx]. Participants received an information sheet detailing the study’s purpose, risks, and data protection measures per Article 14 of the EU GDPR. They provided informed consent, and confidentiality was ensured by prohibiting disclosure of specific companies, individuals, products, or locations. Data were AES-256 encrypted and securely stored, complying with the Data Protection Act 2018 (registration no.: xxxx)
. Following The Menlo Report’s principles, including Respect for Persons and Beneficence, participants had the right to withdraw at any time, with their data excluded if they did. No participants withdrew. By prioritizing transparency and risk minimization, we upheld the highest ethical standards.

\section{Findings and Outcomes}\label{findings}
Our exploration of the integration of AI with smart home devices for privacy protection reveals two main categories: \textbf{Aspirations for AI-Enhanced Privacy}, and \textbf{AI Ethical, Security, and Regulatory Considerations}. Under \textit{Aspirations for AI-Enhanced Privacy}, key themes include empowering passive and vulnerable users with control over their data, real-time privacy alerts, personalized privacy recommendations, and simplified voice-activated privacy management. In the \textit{AI Ethical, Security, and Regulatory Considerations} category, the focus is on ensuring data security through strong encryption and anonymization, compliance with global privacy regulations, transparency in AI decision-making, and minimizing biases while respecting user consent. These themes highlight both the benefits and challenges of integrating AI tools like ChatGPT to enhance privacy protection in smart homes. (The identified categories and themes are summarized in Table-\textcolor{blue}{\ref{Table: Categories and Themes}}).

\subsection{Aspirations for AI-Enhanced Privacy}
This section examines ChatGPT's current privacy protection capabilities in smart homes and highlights users' aspirations for enhancing privacy through AI tools.

\subsubsection{\textbf{Current ChatGPT Capabilities}}

Participants believe AI tools like ChatGPT can help address privacy concerns in smart home devices, including TVs, refrigerators, cameras, security systems, and locks. It offers tailored privacy guidance, such as recommending secure network configurations, implementing WPA3 encryption, isolating devices on guest networks, and adjusting privacy settings to reduce data exposure. ChatGPT also supports privacy-enhancing measures like end-to-end encryption, local data processing, and firewall configurations, enabling users to manage privacy risks proactively. Additionally, it helps users understand regulations like GDPR and CCPA, emphasizing local data storage and anonymized processing. One participant [D02] stated: “ChatGPT can provide privacy guidance and can help improve protection.” While ChatGPT provides comprehensive guidance, professional evaluations remain essential for critical systems like smart security cameras. Future advancements could enhance support for vulnerable user groups, such as children, by offering age-appropriate privacy recommendations, parental control guidance, and protection tailored to their unique risks.

\subsubsection{\textbf{AI For User Empowerment}}
The findings indicate that integrating AI tools into smart home devices empowers users by providing easier access to privacy controls and improving understanding of data management, making privacy protection more accessible and manageable.


\paragraph{\textbf{Privacy Education.}} 

Our participants—users—emphasized that many lack the knowledge to manage privacy in smart home environments, and AI tools like ChatGPT can bridge this gap by offering easy-to-understand guidance on privacy best practices, enabling users to take a proactive role in protecting their data. [PU05] remarked: \textit{“I think ChatGPT can teach me more about smart home privacy than any manual ever could. It will be like having a privacy expert on hand”}. Designers stressed the importance of embedding educational content directly into AI responses to enhance users' understanding and foster trust in AI's potential to protect data. [D03] commented: \textit{“We need to equip ChatGPT and similar tools with tips and educational materials to ensure users are continually learning how to protect themselves”}. Transparency about data usage is crucial, with AI tools able to demystify data practices, offering clear explanations of personal data usage, helping users feel more in control. [PU06] remarked: \textit{“It’s reassuring when ChatGPT tells me exactly how my data is being used. I would feel much more in control”}. Addressing the needs of vulnerable groups like children presents unique challenges. [PU06] stated: \textit{“AI tools must provide age-appropriate, engaging, and easily comprehensible privacy guidance while ensuring children grasp the risks of data sharing in smart home environments”}. Designers suggested parental controls and adaptive learning mechanisms to tailor recommendations to different maturity levels. However, regulators noted that protecting children’s data while delivering personalized education requires strict compliance with child protection regulations (e.g., COPPA~\footnote{\href{https://www.ftc.gov/legal-library/browse/rules/childrens-online-privacy-protection-rule-coppa}{\underline{\textcolor{blue}{Children's Online Privacy Protection Rule\label{COPPA}}}}}) and ethical standards.

\paragraph{\textbf{Power Dynamics and Control Over Data.}}
In smart home contexts, users are increasingly aware of the critical need to maintain control over their personal data as devices continuously collect sensitive information. Integrating AI tools, such as ChatGPT, presents a promising opportunity by providing intuitive, accessible interfaces that enable users to manage their data effectively within the smart home ecosystem. These tools allow users to monitor data collection, understand storage practices, and control access, fostering transparency and trust between users and their devices. By enhancing data management capabilities, AI tools help balance privacy and convenience, empowering users to take proactive steps to protect their personal information while mitigating power imbalances in smart home environments. As [PU01] remarked: \textit{“I want to know exactly what data my smart devices are collecting, and with ChatGPT, I think I will feel like I have that control at my fingertips”}. However, designers argued that while AI integration offers substantial benefits, it also introduces significant challenges for vulnerable user groups, such as children, the elderly, and individuals with disabilities. [D02] argued: \textit{"These users may struggle to navigate AI-driven privacy settings due to cognitive, developmental, or accessibility constraints, increasing the risk of privacy violations. For instance, children may lack the awareness needed to recognize data-sharing risks, while older adults may find AI interfaces unintuitive. To address these challenges, AI-enhanced privacy controls must incorporate inclusive design principles, ensuring equitable access for all household members"}.  

In shared living situations, control over privacy settings often favors admin users or individuals with greater technical expertise, exacerbating power imbalances. This issue is particularly concerning for vulnerable users, who may have limited control over their personal data. To mitigate these risks, device designers are requested to develop user-friendly AI interfaces that simplify interactions with privacy settings, making them accessible even to those with minimal technical knowledge. These designs aim to prevent any single individual from dominating privacy decisions, fostering a more inclusive and balanced approach to privacy management in smart homes. As [D01] explained: \textit{“Our goal is to make the privacy settings so straightforward that users can adjust them without any technical knowledge, just by talking to ChatGPT”}. Despite these efforts, ensuring effective privacy protections for vulnerable groups remains a complex challenge. From their side, regulators added that striking a balance between usability, privacy control, and regulatory compliance (e.g., COPPA for children's data protection) requires ongoing refinement of AI-driven solutions. They also added that future research should explore adaptive learning mechanisms and AI-powered parental controls that dynamically adjust recommendations based on user maturity levels. Addressing these challenges is essential to making AI-enhanced privacy tools truly inclusive and effective for all users in smart home environments.

\paragraph{\textbf{Real-Time Privacy Alerts.}}
Findings show that real-time privacy alerts are essential in smart homes for maintaining user trust and securing data. Users value immediate notifications from AI tools like ChatGPT, enabling them to take timely action and manage privacy actively. This constant awareness reinforces their sense of control. [PU02] said: \textit{``If there's a privacy risk, I want to be the first to know, and I think ChatGPT makes that happen almost instantly''}. AI developers stress the importance of training AI tools to detect and respond to privacy threats swiftly. The goal is for tools to be both reactive and proactive, flagging suspicious patterns before escalation, thus helping users maintain continuous privacy monitoring without manual oversight. [AI01] said: \textit{``We need to make tools like ChatGPT proactive in privacy protection, notifying users immediately when something suspicious occurs''}.

\subsubsection{\textbf{AI For Personalization and Adaptation}}
Findings reveal that AI tools' potential to adapt to individual preferences is a key advancement in privacy protection. By understanding user behaviors and contexts, these tools provide tailored privacy recommendations, ensuring a more personalized experience.

\paragraph{\textbf{Context-Aware Privacy Recommendations.}}
Participants highlighted that one of ChatGPT's standout features is its ability to provide context-aware privacy recommendations, adapting its advice to specific situations like being at home, traveling, or engaged in certain activities. This ensures privacy settings are both effective and practical, aligning with the user’s environment and lifestyle. As [PU03] noted: \textit{“The fact that AI tools like ChatGPT can know my routines and suggest privacy settings that fit my lifestyle is incredibly helpful.”} Such tailored recommendations enhance the user experience, offering practical solutions for varying scenarios. Experts stress the importance of advancing AI tools to deliver more context-specific recommendations, especially in smart home environments. The goal is to provide relevant privacy advice for unique situations, such as managing multiple devices or adjusting settings based on daily interactions with technology. This personalized approach makes privacy management seamless and intuitive, empowering users to implement recommendations and maintain control over their data across devices. As [D02] expressed: \textit{“We need to design AI tools to be more intuitive, so they can suggest privacy settings based on users’ activities.”} This vision underscores the need for aligning AI capabilities with user-centric designs to foster trust and usability in smart home ecosystems.

\paragraph{\textbf{Seamless Integration with Smart Devices.}}


AI developers emphasize that effective privacy protection in smart homes requires seamless integration of AI tools with devices. GenAI provides personalized, context-aware privacy recommendations, enhancing user experience without compromising functionality. While ConvAI aids interaction, GenAI is more effective in balancing privacy with convenience. This integration ensures privacy protection does not disrupt smart device usability. [AI04] said: \textit{Users don’t want privacy protection to cost convenience. GenAI tools need to work seamlessly with smart devices while offering personalized privacy recommendations''}. Developers stress creating seamless interactions with ChatGPT so users benefit from enhanced privacy without noticing the technology. This unobtrusive integration allows users to enjoy smart devices securely. [AI02] said: \textit{The key is making sure ChatGPT integrates so well with devices that users don’t notice it but still feel protected''}.

\subsubsection{\textbf{AI For Ease of Use and Accessibility}}
Effective privacy protection must be easy to use and accessible to all, with AI tools simplifying privacy management for users of all technical proficiencies.

\paragraph{\textbf{Simplified Privacy Controls.}}
Participants --admins-- argued that users often find privacy settings on smart devices to be confusing or overly complex, particularly in the context of smart homes. They added that AI tools can simplify these controls, making them more intuitive and easier to manage. By providing straightforward options and clear explanations, AI tools ensure that users can quickly and easily adjust their privacy settings across multiple devices without feeling overwhelmed. [AD01] said: \textit{``Privacy settings should be as easy as flipping a switch''}. Designers emphasized the need to reduce the complexity of interacting with privacy settings through AI tools in smart homes, ensuring that these settings are accessible to all users, regardless of their technical background. This approach aims to make privacy management within the smart home environment as seamless and user-friendly as possible. [D04] stated: \textit{``We need to eliminate the jargon and ensure that privacy controls of AI tools are intuitive for everyone''}.  

However, ensuring effective privacy management for vulnerable user groups, such as children, the elderly, and individuals with disabilities, presents unique challenges. Regulators highlighted that AI-driven privacy controls must be designed with inclusivity in mind, incorporating adaptive interfaces and parental controls to protect children's data while maintaining ease of use. They also stressed the importance of compliance with child protection regulations (e.g., COPPA~\footref{COPPA}) to ensure AI tools protect children's privacy effectively.

\paragraph{\textbf{Voice-Activated Privacy Management.}}
Participants --admins-- discussed how voice commands provide a convenient way for users to manage privacy settings in the smart home. They noted that AI tools can improve this by responding to natural language queries, allowing hands-free privacy control across devices. This feature enhances accessibility and convenience, enabling users to maintain privacy without disrupting daily routines. [AD02] said: \textit{Being able to manage my privacy with just my voice is a game-changer, and I think AI can make it feel natural''}. AI developers stressed the need for accurate voice-activated controls that reliably execute user commands within smart home systems. [AI04] said: \textit{Voice activation is crucial for accessibility. We need to ensure that ChatGPT accurately understands and executes commands every time''}.

\subsection{Ethical and Regulatory Considerations}
Integrating AI with smart home devices offers privacy benefits but raises ethical, security, and regulatory concerns, highlighting the need for legal data protection, and ethical AI use.


\subsubsection{\textbf{Transparency and Accountability}}

User data is vital for AI functions but must be securely protected in smart home systems. Designers emphasize robust encryption for safeguarding data both at rest and in transit, ensuring sensitive information, such as usage patterns, remains secure and maintains user trust. Encryption prevents unauthorized access, offering peace of mind about privacy protection. [D01] noted: \textit{Knowing that my data is encrypted gives me peace of mind. It’s good to see AI tools prioritizing this''}. Participants also highlighted data anonymization to eliminate personally identifiable information, ensuring compromised data cannot be traced back to individuals. Combining anonymization with data minimization helps protect privacy while maintaining functionality, which is essential for effective smart home systems. [D01] added: \textit{By minimizing the data we collect and anonymizing what we do, we’re striking a balance between privacy and functionality''}. Users and regulators demand transparency in AI decision-making and accountability for breaches. Participants stressed understanding how AI tools, like ChatGPT, manage privacy settings, as transparency fosters trust. Clear explanations demystify AI processes, especially for managing multiple devices. [AD03] remarked: \textit{I want to know why ChatGPT recommends certain privacy settings. Transparency is crucial''}. Regulators and developers agree decision-making must be clear to ensure user control and accountability. [AI04] noted: \textit{Explainability is a priority, especially for managing privacy across devices''}. In case of breaches, clear accountability with response protocols is vital. [R03] stated: \textit{If something goes wrong, I need to know who’s responsible''}.

\subsubsection{\textbf{Compliance with Regulations and AI Ethical Use}}
Participants emphasized the need for AI tools like ChatGPT to comply with privacy regulations such as GDPR and CCPA to build trust and protect users. Designers highlighted the importance of adhering to international privacy laws, ensuring ethical data handling across jurisdictions. [D03] noted: \textit{“I need to know that ChatGPT is respecting privacy laws wherever I am.”} Regulators stressed collaboration between AI developers and policymakers to ensure compliance, addressing concerns about data residency and sovereignty. Developers are working on adaptable systems to meet local legal requirements, with [R01] stating: \textit{“We want to know where the data is kept and that it’s not leaving the country without permission.”} Ethical concerns, such as fairness, bias minimization, and user autonomy, remain critical, with users and regulators warning that AI biases could lead to unfair outcomes. Continuous monitoring is necessary to ensure impartiality, as [PU04] mentioned: \textit{“I need to trust that ChatGPT isn’t biased in handling my privacy.”} Participants also stressed the importance of securing explicit user consent before making privacy-related decisions, with [AD04] emphasizing: \textit{“I want to ensure ChatGPT only acts on my behalf with my explicit approval.”}




\section{Discussion}

\subsection{Summary of Findings}
This study explores integrating ChatGPT and similar AI tools with smart home devices to enhance privacy protection. We engaged users, designers, AI developers, and regulators to gather their perspectives, organizing the findings into two main areas: \textit{Aspirations for AI-Enhanced Privacy} and \textit{AI Ethical, Security, and Regulatory Considerations}. 

The first category emphasized AI tools' potential to give users greater control over privacy, with participants valuing features like real-time alerts, intuitive data management, and personalized recommendations. ChatGPT’s role in educating users about privacy risks and its accessibility through voice commands enhanced the balance between convenience and security. Addressing the needs of vulnerable groups, such as children, presented challenges, with participants stressing the importance of age-appropriate, accessible, and easy-to-use interfaces. Designers highlighted the need for adaptive learning mechanisms and parental controls for users with varying digital literacy. The second category focused on data security through encryption, anonymization, global privacy law compliance, and transparency in AI decision-making. Clear accountability for breaches, minimizing biases, and ensuring informed consent were emphasized for ethical AI integration. Regulators stressed adherence to child data protection regulations (e.g., COPPA) and ethical safeguards when designing privacy solutions for vulnerable populations.

Integrating AI tools like ChatGPT with smart home devices can enhance privacy protection, but usability, accessibility, and regulatory compliance are key challenges. The following sections discuss recommendations for improving privacy strategies.

\subsection{Power Dynamics and User Empowerment}
Integrating AI tools with smart home devices has the potential to revolutionize privacy management by addressing the critical intersection of privacy concerns and household power dynamics. Building on prior research~\cite{oladoyinbo_exploring_2024,tiwari_inbuilt_2023,azar_home_2023}, this study explores how AI capabilities—such as real-time alerts, voice-activated commands, and adaptive personalization—can empower users to proactively manage their privacy settings. By enabling dynamic, user-tailored adjustments, AI enhances autonomy and addresses privacy concerns at an individual level. However, significant challenges persist, particularly in shared smart home environments, where power imbalances among household members—such as technically proficient users dominating privacy settings—can marginalize vulnerable users, including children, older adults, and individuals with limited digital literacy.

The findings emphasize the transformative potential of GenAI in addressing these challenges. Acting as a neutral intermediary, GenAI can democratize privacy control by simplifying complex settings, enhancing accessibility, and mitigating power imbalances. This aligns with prior research~\cite{albayaydh_examining_2023,bernd_bystanders_2019,bernd_balancing_2022} that underscores the importance of user education and awareness in fostering equitable participation. GenAI’s natural language understanding and adaptability allow it to educate less proficient users, enhance awareness, and facilitate collaborative decision-making. For instance, GenAI could offer interactive tutorials or real-time guidance tailored to specific privacy scenarios, helping passive and vulnerable users make informed choices. These features bridge knowledge gaps and promote inclusivity, but careful design is crucial to prevent misuse, such as skilled users exploiting their expertise to override shared decisions. Embedding features that support transparent, collaborative privacy management ensures that all household members feel empowered, fostering smart home ecosystems built on inclusivity and fairness.

\subsection{Education and User Awareness}
Findings indicate that AI tools can play a critical role in enhancing user education about privacy risks and best practices, particularly within smart home settings. By providing accessible explanations and tailored guidance on privacy concerns, AI tools help users become more informed and proactive in managing their privacy across interconnected devices. This educational role is especially beneficial in smart homes, where the complexity of managing privacy across multiple devices can be overwhelming, particularly for vulnerable users, and individuals with limited digital literacy. By increasing user awareness and engagement with privacy settings, AI tools empower users to make better-informed decisions about their personal data. The study’s findings align with prior research~\cite{albayaydh_examining_2023,albayaydh_exploring_2022,bernd_balancing_2022}, which emphasizes the importance of educational support in fostering user autonomy in privacy management within digital contexts.

However, the challenge lies in ensuring that AI tools provide clear, accurate, and accessible information for all users, including those with limited technological expertise. Vulnerable users may struggle with complex privacy settings, requiring simplified, interactive guidance tailored to various levels of digital literacy. Findings suggest that user-friendly explanations and support materials are crucial for maximizing the educational impact of AI tools in smart homes, where users interact with multiple devices. Prior research~\cite{manheim_artificial_2019,furey_can_2019,zhu_applying_2019} emphasizes the need for simplicity and clarity in privacy communication. The study highlights how AI-driven, context-aware explanations—delivered through conversational interfaces or interactive tutorials—can improve user comprehension and encourage more engagement with privacy settings, making privacy education more inclusive and effective.

\subsection{Ethical and Security Considerations}
The study highlights that integrating GenAI tools with smart home devices enhances privacy and security but also presents ethical, legal, and security challenges. These include detecting potential abuse, such as partner violence or landlord overreach, through behavioral patterns and contextual clues. Ethical concerns, such as cultural biases, false positives, and unintended consequences of automated decision-making, may arise due to varying privacy norms and data expectations. The study stresses the need for culturally sensitive AI tools, transparency, and legal frameworks, especially concerning privacy and domestic violence laws. A nuanced approach is required to balance privacy with legal obligations, particularly regarding AI-generated insights shared with authorities. Ongoing collaboration between AI developers, legal experts, and ethicists is essential to minimize biases, refine abuse detection accuracy, and protect privacy~\cite{kraemer_informing_2019, albayaydh_innovative_2023}.

Furthermore, transparency in AI decision-making is critical for fostering user trust in smart home systems. Users need to understand how AI tools manage privacy settings and data, with clear explanations of AI processes to demystify decision-making. This supports user engagement and trust, as shown in prior studies on explainability~\cite{albayaydh_examining_2023, albayaydh_innovative_2023}. Establishing transparent, responsive complaint-handling procedures helps address privacy issues and strengthens the system's credibility. This proactive approach, informed by AI transparency research~\cite{ramya_pillars_2025, leenes_chapter_2018}, is vital for ensuring users feel secure and in control in an evolving data-driven environment.

\subsection{Recommendations}

\subsubsection{\textbf{For Users and Social Entities}}
The study underscores the critical role of user awareness in managing privacy risks in smart homes, reinforcing previous research on the impact of education in privacy management~\cite{albayaydh_examining_2023,bernd_balancing_2022}. Two key recommendations are proposed:

\textbf{Enhance Awareness and Engage with AI Tools:} Users should educate themselves and their dependents, including vulnerable groups such as children and elderly individuals, about privacy risks and best practices for smart home devices. Regularly updating settings to address emerging threats is essential. AI tools like ChatGPT can play a key role in making privacy management accessible to all users by providing clear, simple explanations. Social organizations should prioritize inclusive programs that educate diverse communities on privacy, ensuring everyone, regardless of technical skill, has the information needed to make informed decisions. Actively utilizing AI-driven privacy features, such as those available through ChatGPT, strengthens user control and security within the smart home. For vulnerable users, these tools can be designed with simplicity in mind, offering real-time alerts, straightforward privacy preferences, and intuitive interfaces. By personalizing privacy strategies, AI tools can help users effectively manage privacy risks and adjust settings across interconnected devices.

\textbf{Support Ethical AI Development:} The study emphasizes that social organizations play a vital role in advocating for ethical AI practices in smart home systems. Supporting efforts to reduce AI biases and ensuring robust privacy protections are key to AI’s ethical integration into daily life. Additionally, developers must ensure that AI privacy tools are usable for all, including vulnerable groups, by offering adaptive learning features and providing necessary support for users who may have difficulties navigating complex settings. Encouraging research and policy initiatives that promote responsible AI use will foster systems that prioritize user privacy and cater to diverse needs.

\subsubsection{\textbf{For Smart Device Designers and Manufacturers}}
The study’s findings underscore two critical recommendations for enhancing privacy management in smart homes:

\textbf{Prioritize User-Centric Design:} Findings reveal that intuitive, user-friendly interfaces are fundamental for effective privacy management across smart home devices. Designers should focus on creating adaptive, context-aware interfaces that simplify the management of privacy settings, even for users with minimal technical expertise. This entails seamless integration of AI tools that dynamically adjust privacy settings based on user behavior and preferences, accommodating diverse household roles and technical proficiencies. Such systems could employ features like guided walkthroughs, natural language interaction, and visual aids to make privacy controls more accessible and engaging. By addressing varying expertise levels, designers can empower all household members—passive and vulnerable users—to take ownership of their privacy. Building on prior studies~\cite{westin_opt_2019,chhetri_eliciting_2019}, this user-centered approach fosters autonomy and inclusivity, ensuring no individual is marginalized in privacy decision-making.

\textbf{Implement Robust Security Measures:} The findings underscore the need to embed advanced security frameworks within smart home ecosystems. Manufacturers should implement cutting-edge encryption, anonymization, and edge computing to protect user data while minimizing risks associated with centralized storage and transfer. Beyond compliance with regulations like GDPR and CCPA, proactive measures such as AI-driven threat detection, real-time security notifications, regular audits, and automatic updates are vital for maintaining trust. User-facing tools, like personalized security dashboards, can further empower users to monitor and address potential threats, promoting proactive privacy management. This holistic approach blends user empowerment with robust security practices, fostering innovative and ethically aligned smart home ecosystems that meet the expectations of privacy-conscious users.

\subsubsection{\textbf{For AI Developers}}
To improve privacy, we offer two key recommendations:

\textbf{Enhance AI Accuracy and Personalization:} Personalized, AI-driven privacy recommendations enhance user trust and satisfaction. Developers should refine algorithms to provide tailored privacy guidance, considering diverse demographics, cultural practices, and household dynamics. Regular updates informed by user feedback are essential to address emerging challenges. AI tools should also include real-time alerts for privacy violations, empowering users to act and control their data. Contextual awareness ensures privacy recommendations remain relevant and actionable. This approach aligns with prior research~\cite{chhetri_user-centric_2022, olaoye_enhancing_2024}, emphasizing personalization, particularly for vulnerable users with limited technical expertise.

\textbf{Address Ethical Concerns:} The study emphasizes tackling ethical issues such as bias, consent, and transparency in AI development. Developers must minimize biases to ensure fairness and prevent marginalization. Transparent decision-making helps users understand how privacy recommendations are generated. Informed consent should guide AI interactions with accessible protocols for data handling. Offering clear data usage explanations and customizable privacy settings fosters trust and engagement. Addressing these ethical aspects ensures AI tools meet user expectations and regulatory standards, promoting robust privacy frameworks, especially for vulnerable groups who may struggle with AI systems.

\subsubsection{\textbf{For Regulators}}
To support smart home privacy, we offer two suggestions:

\textbf{Strengthen Privacy Regulations:} Findings highlight the need for up-to-date privacy regulations that address the unique challenges AI and smart home technologies introduce. Regulators should consider updating privacy laws to encompass AI-driven systems within smart homes, ensuring that user data is adequately protected against emerging risks. The study also points to the importance of harmonizing international privacy standards, enabling tech companies to comply with both local and global laws more effectively. Additionally, these regulations should ensure that privacy tools are designed to be accessible to vulnerable groups, such as children, elderly individuals, or those with limited technical expertise, fostering inclusivity.

\textbf{Promote Transparency and Accountability:} The study highlights that transparent AI practices and clear accountability are crucial for building user confidence in AI-integrated smart home devices. Governments should promote transparency in AI decision-making by establishing guidelines on data handling and protocols for addressing privacy violations and breaches. This ensures accountability and maintains user trust. Encouraging transparency within smart home contexts boosts public awareness, strengthens users' sense of security, and ensures that privacy controls are easy to understand and manage, particularly for vulnerable users who may struggle with complex technical settings.


\section{Conclusion}
The integration of AI tools with smart home devices presents a transformative opportunity to enhance user privacy management. Our findings show that AI tools can significantly empower users, including vulnerable groups such as children, elderly individuals, and those with limited technical expertise, by providing intuitive interfaces to manage privacy settings, fostering trust and control over data collection, storage, and access. However, the distribution of privacy control among household members can create imbalances, underscoring the need for equitable access to privacy management features.

While AI tools offer personalization and education on privacy risks, challenges remain in addressing power dynamics and ensuring privacy control effectiveness. Accurate interpretation and execution of commands are crucial for maintaining user trust, especially for vulnerable users who may find privacy settings complex. Ethical considerations, such as robust security measures to prevent data breaches, are also essential. Our recommendations focus on enhancing user awareness and engagement with GenAI-enabled smart devices, prioritizing user-centric design that accommodates varying levels of technical expertise, and implementing robust security. Additionally, improving AI accuracy, minimizing biases, and strengthening privacy regulations will contribute to more secure and equitable privacy management, making it accessible to all users in the evolving smart home landscape.

\enlargethispage{10pt}


\section*{Acknowledgments}
The authors sincerely thank the staff of our institution for their support and resources throughout this study. We also appreciate the valuable contributions of all participants, whose insights and efforts were instrumental in shaping our findings.


\bibliographystyle{plain}
\bibliography{cited_only}

\begin{thebibliography}{100}

\bibitem{acquisti_nudges_2018}
Alessandro Acquisti, Idris Adjerid, Rebecca Balebako, Laura Brandimarte, Lorrie~Faith Cranor, Saranga Komanduri, Pedro~Giovanni Leon, Norman Sadeh, Florian Schaub, Manya Sleeper, Yang Wang, and Shomir Wilson.
\newblock Nudges for {Privacy} and {Security}: {Understanding} and {Assisting} {Users}’ {Choices} {Online}.
\newblock {\em ACM Computing Surveys}, 50(3):1--41, May 2018.

\bibitem{ahmad_tangible_2020}
Imtiaz Ahmad, Rosta Farzan, Apu Kapadia, and Adam~J. Lee.
\newblock Tangible {Privacy}: {Towards} {User}-{Centric} {Sensor} {Designs} for {Bystander} {Privacy}.
\newblock {\em Proceedings of the ACM on Human-Computer Interaction}, 4(CSCW2):1--28, October 2020.

\bibitem{ahmed_understanding_2017}
Tousif Ahmed, Roberto Hoyle, Patrick Shaffer, Kay Connelly, David Crandall, and Apu Kapadia.
\newblock Understanding {Physical} {Safety}, {Security}, and {Privacy} {Concerns} of {People} with {Visual} {Impairments}.
\newblock {\em IEEE Internet Computing}, pages 1--1, 2017.

\bibitem{al-alami_vulnerability_2017}
Haneen Al-Alami, Ali Hadi, and Hussein Al-Bahadili.
\newblock Vulnerability scanning of {IoT} devices in {Jordan} using {Shodan}.
\newblock pages 1--6, December 2017.

\bibitem{albayaydh_examining_2023}
Wael Albayaydh and Ivan Flechais.
\newblock Examining {Power} {Dynamics} and {User} {Privacy} in {Smart} {Technology} {Use} {Among} {Jordanian} {Households}.
\newblock pages 4643--4659, 2023.

\bibitem{albayaydh_innovative_2023}
Wael Albayaydh and Ivan Flechais.
\newblock "{Innovative} {Technologies} or {Invasive} {Technologies}?": {Exploring} {Design} {Challenges} of {Privacy} {Protection} {With} {Smart} {Home} in {Jordan}.
\newblock {\em Proc. ACM Hum.-Comput. Interact., Vol. 1, No. CSCW}, 1(Proc. ACM Hum.-Comput. Interact., Vol. 1, No. CSCW), 2023.

\bibitem{albayaydh_co-designing_2024}
Wael Albayaydh and Ivan Flechais.
\newblock Co-{Designing} a {Mobile} {App} for {Bystander} {Privacy} {Protection} in {Jordanian} {Smart} {Homes}: {A} {Step} {Towards} {Addressing} a {Complex} {Privacy} {Landscape}.
\newblock {\em USENIX Security}, 2024.

\bibitem{albayaydh_exploring_2022}
Wael~S Albayaydh and Ivan Flechais.
\newblock Exploring {Bystanders}’ {Privacy} {Concerns} with {Smart} {Homes} in {Jordan}.
\newblock In {\em {CHI} {Conference} on {Human} {Factors} in {Computing} {Systems}}, pages 1--24, New Orleans LA USA, April 2022. ACM.

\bibitem{ali_cyber_2018}
Bako Ali and Ali~Ismail Awad.
\newblock Cyber and {Physical} {Security} {Vulnerability} {Assessment} for {IoT}-{Based} {Smart} {Homes}.
\newblock {\em Sensors}, 18(3):817, March 2018.

\bibitem{apthorpe_you_2022}
Noah Apthorpe, Pardis Emami-Naeini, Arunesh Mathur, Marshini Chetty, and Nick Feamster.
\newblock You, {Me}, and {IoT}: {How} {Internet}-{Connected} {Consumer} {Devices} {Affect} {Interpersonal} {Relationships}.
\newblock {\em ACM Transactions on Internet of Things}, page 3539737, June 2022.

\bibitem{azar_home_2023}
Joseph Azar, Teressa Khoury, Abdallah Makhoul, and Raphael Couturier.
\newblock Home {Automation} {System} with {IoT} {Stack} and {ChatGPT} for {People} with {Reduced} {Mobility}.
\newblock In {\em 2023 {IEEE} 4th {International} {Multidisciplinary} {Conference} on {Engineering} {Technology} ({IMCET})}, pages 44--49, December 2023.

\bibitem{badii_smart_2020}
Claudio Badii, Pierfrancesco Bellini, Angelo Difino, and Paolo Nesi.
\newblock Smart {City} {IoT} {Platform} {Respecting} {GDPR} {Privacy} and {Security} {Aspects}.
\newblock {\em IEEE Access}, 8:23601--23623, 2020.

\bibitem{ball_workplace_2010}
Kirstie Ball.
\newblock Workplace surveillance: an overview. {Kirstie} {Ball} (2010) {Workplace} surveillance: an overview, {Labor} {History}.
\newblock {\em Labor History}, 51(1):87--106, February 2010.

\bibitem{bastos_gdpr_2018}
Daniel Bastos, Fabio Giubilo, Mark Shackleton, and Fadi El-Mousa.
\newblock {\em {GDPR} {Privacy} {Implications} for the {Internet} of {Things}}.
\newblock December 2018.

\bibitem{baxter_understanding_2015}
Kathy Baxter, Catherine Courage, and Kelly Caine.
\newblock {\em Understanding {Your} {Users}: {A} {Practical} {Guide} to {User} {Research} {Methods}}.
\newblock Morgan Kaufmann, May 2015.

\bibitem{bennett_european_2018}
Colin~J. Bennett.
\newblock The {European} {General} {Data} {Protection} {Regulation}: {An} instrument for the globalization of privacy standards?
\newblock {\em Information Polity}, 23(2):239--246, June 2018.

\bibitem{benthall_contexts_2019}
Sebastian Benthall and Bruce~D Haynes.
\newblock Contexts are {Political}: {Field} {Theory} and {Privacy}. {Association} for {Computing} {Machinery}. {ACM}.
\newblock page~3, 2019.

\bibitem{bernd_balancing_2022}
Julia Bernd, Ruba Abu-Salma, Junghyun Choy, and Alisa Frik.
\newblock Balancing {Power} {Dynamics} in {Smart} {Homes}: {Nannies}' {Perspectives} on {How} {Cameras} {Reflect} and {Affect} {Relationships}.
\newblock page~21, 2022.

\bibitem{bernd_bystanders_2019}
Julia Bernd, Ruba Abu-Salma, and Alisa Frik.
\newblock Bystanders’ {Privacy}: {The} {Perspectives} of {Nannies} on {Smart} {Home} {Surveillance},.
\newblock page~14, 2019.

\bibitem{brand_survey_2020}
Denys Brand, Florence~D. DiGennaro~Reed, Mariah~D. Morley, Tyler~G. Erath, and Matthew~D. Novak.
\newblock A {Survey} {Assessing} {Privacy} {Concerns} of {Smart}-{Home} {Services} {Provided} to {Individuals} with {Disabilities}.
\newblock {\em Behavior Analysis in Practice}, 13(1):11--21, March 2020.

\bibitem{bubeck_sparks_2023}
Sébastien Bubeck, Varun Chandrasekaran, Ronen Eldan, Johannes Gehrke, Eric Horvitz, Ece Kamar, Peter Lee, Yin~Tat Lee, Yuanzhi Li, Scott Lundberg, Harsha Nori, Hamid Palangi, Marco~Tulio Ribeiro, and Yi~Zhang.
\newblock Sparks of {Artificial} {General} {Intelligence}: {Early} experiments with {GPT}-4, March 2023.

\bibitem{burke_contemporary_2006}
Peter~J. Burke, editor.
\newblock {\em Contemporary social psychological theories . {Center} for {Internet} and {Society}, {Stanford} {Law} {School} {Andreas} {Katsanevas}, {Department} of {Communication}, {Stanford} {University}}.
\newblock Stanford Social Sciences, Stanford, Calif, 2006.

\bibitem{calder_eu_2020}
Alan Calder.
\newblock {\em {EU} {General} {Data} {Protection} {Regulation} ({GDPR}) – {An} implementation and compliance guide, fourth edition}.
\newblock 2020.

\bibitem{cao_new_2023}
XinXia Cao.
\newblock A new era of intelligent interaction: {Opportunities} and challenges brought by {ChatGPT}, July 2023.

\bibitem{chaudhuri_internet_2016}
Abhik Chaudhuri.
\newblock Internet of things data protection and privacy in the era of the {General} {Data} {Protection} {Regulation}.
\newblock {\em Journal of Data Protection \& Privacy}, 1(1):64--75, December 2016.

\bibitem{chhetri_user-centric_2022}
Chola Chhetri and Vivian Genaro~Motti.
\newblock User-{Centric} {Privacy} {Controls} for {Smart} {Homes}.
\newblock {\em Proceedings of the ACM on Human-Computer Interaction}, 6(CSCW2):1--36, November 2022.

\bibitem{chhetri_eliciting_2019}
Chola Chhetri and Vivian~Genaro Motti.
\newblock Eliciting {Privacy} {Concerns} for {Smart} {Home} {Devices} from a {User} {Centered} {Perspective}.
\newblock In Natalie~Greene Taylor, Caitlin Christian-Lamb, Michelle~H. Martin, and Bonnie Nardi, editors, {\em Information in {Contemporary} {Society}}, Lecture {Notes} in {Computer} {Science}, pages 91--101, Cham, 2019. Springer International Publishing.

\bibitem{collingridge_quality_2008}
Dave~S. Collingridge and Edwin~E. Gantt.
\newblock The {Quality} of {Qualitative} {Research}.
\newblock {\em American Journal of Medical Quality}, 23(5):389--395, September 2008.

\bibitem{corbin_basics_2014}
Juliet Corbin and Anselm Strauss.
\newblock {\em Basics of {Qualitative} {Research}: {Techniques} and {Procedures} for {Developing} {Grounded} {Theory}}.
\newblock SAGE Publications, November 2014.

\bibitem{deceptive-patterns_deceptive_2020}
deceptive patterns.
\newblock Deceptive {Patterns} - {Home}, 2020.

\bibitem{dell_yours_2012}
Nicola Dell, Vidya Vaidyanathan, Indrani Medhi, Edward Cutrell, and William Thies.
\newblock "{Yours} is better!": participant response bias in {HCI}.
\newblock pages 1321--1330, Austin Texas USA, May 2012. ACM.

\bibitem{dhingra_default_2012}
Nikhil Dhingra, Zach Gorn, Andrew Kener, and Jason Dana.
\newblock The default pull: {An} experimental demonstration of subtle default effects on preferences.
\newblock {\em Judgment and Decision Making}, 7(1):69--76, January 2012.

\bibitem{dreyfus_five-stage_1980}
Stuart~E Dreyfus.
\newblock A {Five}-{Stage} {Model} of the {Mental} {Activities} {Involved} in {Directed} {Skill} {Acquisition}.
\newblock Technical report, February 1980.

\bibitem{eckhoff_privacy_2018}
David Eckhoff and Isabel Wagner.
\newblock Privacy in the {Smart} {City}—{Applications}, {Technologies}, {Challenges}, and {Solutions}.
\newblock 20(1):489--516, 2018.

\bibitem{emami-naeini_exploring_2019}
Pardis Emami-Naeini, Henry Dixon, Yuvraj Agarwal, and Lorrie~Faith Cranor.
\newblock Exploring {How} {Privacy} and {Security} {Factor} into {IoT} {Device} {Purchase} {Behavior}.
\newblock In {\em Proceedings of the 2019 {CHI} {Conference} on {Human} {Factors} in {Computing} {Systems}}, pages 1--12, Glasgow Scotland Uk, May 2019. ACM.

\bibitem{fang_llm_2024}
Richard Fang, Rohan Bindu, Akul Gupta, Qiusi Zhan, and Daniel Kang.
\newblock {LLM} {Agents} can {Autonomously} {Hack} {Websites}, February 2024.

\bibitem{farke_are_2021}
Florian~M Farke, David~G Balash, Maximilian Golla, Markus Duermuth, and Adam~J Aviv.
\newblock Are {Privacy} {Dashboards} {Good} for {End} {Users}? {Evaluating} {User} {Perceptions} and {Reactions} to {Google}'s {My} {Activity}.
\newblock 2021.

\bibitem{felfernig_overview_2019}
Alexander Felfernig, Seda Polat-Erdeniz, Christoph Uran, Stefan Reiterer, Muesluem Atas, Thi Ngoc~Trang Tran, Paolo Azzoni, Csaba Kiraly, and Koustabh Dolui.
\newblock An overview of recommender systems in the internet of things.
\newblock {\em Journal of Intelligent Information Systems}, 52(2):285--309, April 2019.

\bibitem{feuerriegel_generative_2024}
Stefan Feuerriegel, Jochen Hartmann, Christian Janiesch, and Patrick Zschech.
\newblock Generative {AI}.
\newblock {\em Business \& Information Systems Engineering}, 66(1):111--126, February 2024.

\bibitem{forlizzi_service_2006}
Jodi Forlizzi and Carl DiSalvo.
\newblock Service {Robots} in the {Domestic} {Environment}: {A} {Study} of the {Roomba} {Vacuum} in the {Home}.
\newblock 2006.

\bibitem{freed_conversational_2021}
Andrew Freed.
\newblock {\em Conversational {AI}. {Design}, develop, and deploy human-like {AI} solutions that chat with your customers, solve their problems}.
\newblock 2021.

\bibitem{furey_can_2019}
Eoghan Furey and Juanita Blue.
\newblock Can {I} {Trust} {Her}? {Intelligent} {Personal} {Assistants} and {GDPR}.
\newblock In {\em 2019 {International} {Symposium} on {Networks}, {Computers} and {Communications} ({ISNCC})}, pages 1--6, June 2019.

\bibitem{geeng_whos_2019}
Christine Geeng and Franziska Roesner.
\newblock Who's {In} {Control}?: {Interactions} {In} {Multi}-{User} {Smart} {Homes}.
\newblock In {\em Proceedings of the 2019 {CHI} {Conference} on {Human} {Factors} in {Computing} {Systems}}, pages 1--13, Glasgow Scotland Uk, May 2019. ACM.

\bibitem{gerber_foxit_2018}
Nina Gerber, Paul Gerber, Hannah Drews, Elisa Kirchner, Noah Schlegel, Tim Schmidt, and Lena Scholz.
\newblock {FoxIT}: enhancing mobile users' privacy behavior by increasing knowledge and awareness.
\newblock In {\em Proceedings of the 7th {Workshop} on {Socio}-{Technical} {Aspects} in {Security} and {Trust}}, pages 53--63, Orlando Florida USA, December 2018. ACM.

\bibitem{ghiglieri_exploring_2017}
Marco Ghiglieri, Melanie Volkamer, and Karen Renaud.
\newblock Exploring {Consumers}’ {Attitudes} of {Smart} {TV} {Related} {Privacy} {Risks}.
\newblock Lecture {Notes} in {Computer} {Science}, pages 656--674, Cham, 2017. Springer International Publishing.

\bibitem{glaser_discovery_1967}
Barney~G. Glaser and Anselm~L. Strauss.
\newblock {\em The {Discovery} of {Grounded} {Theory}: {Strategies} for {Qualitative} {Research}}.
\newblock Aldine Transaction, 1967.

\bibitem{goodman_snowball_1961}
Leo~A. Goodman.
\newblock Snowball {Sampling}.
\newblock {\em The Annals of Mathematical Statistics}, 32(1):148--170, 1961.

\bibitem{guest_how_2006}
Greg Guest, Arwen Bunce, and Laura Johnson.
\newblock How {Many} {Interviews} {Are} {Enough}?: {An} {Experiment} with {Data} {Saturation} and {Variability}.
\newblock {\em Field Methods}, 18(1):59--82, February 2006.

\bibitem{gupta_chatgpt_2023}
Maanak Gupta, CharanKumar Akiri, Kshitiz Aryal, Eli Parker, and Lopamudra Praharaj.
\newblock From {ChatGPT} to {ThreatGPT}: {Impact} of {Generative} {AI} in {Cybersecurity} and {Privacy}, July 2023.

\bibitem{habib_evaluating_2022}
Hana Habib and Lorrie~Faith Cranor.
\newblock Evaluating the {Usability} of {Privacy} {Choice} {Mechanisms}.
\newblock 2022.

\bibitem{hsu_awareness_2020}
Silas Hsu, Kristen Vaccaro, Yin Yue, Aimee Rickman, and Karrie Karahalios.
\newblock Awareness, {Navigation}, and {Use} of {Feed} {Control} {Settings} {Online}.
\newblock In {\em Proceedings of the 2020 {CHI} {Conference} on {Human} {Factors} in {Computing} {Systems}}, pages 1--13, Honolulu HI USA, April 2020. ACM.

\bibitem{ioannidou_general_2021}
Irene Ioannidou and Nicolas Sklavos.
\newblock On {General} {Data} {Protection} {Regulation} {Vulnerabilities} and {Privacy} {Issues}, for {Wearable} {Devices} and {Fitness} {Tracking} {Applications}.
\newblock {\em Cryptography}, 5(4):29, October 2021.

\bibitem{j_kraemer_exploring_2019}
Martin J~Kraemer, Ivan Flechais, and Helena Webb.
\newblock Exploring {Communal} {Technology} {Use} in the {Home}.
\newblock pages 1--8, Nottingham United Kingdom, November 2019. ACM.

\bibitem{jakobi_catches_2017}
Timo Jakobi, Corinna Ogonowski, Nico Castelli, Gunnar Stevens, and Volker Wulf.
\newblock The {Catch}(es) with {Smart} {Home}: {Experiences} of a {Living} {Lab} {Field} {Study}.
\newblock In {\em Proceedings of the 2017 {CHI} {Conference} on {Human} {Factors} in {Computing} {Systems}}, pages 1620--1633, Denver Colorado USA, May 2017. ACM.

\bibitem{johnson_beyond_2020}
Mark Johnson, Maggy Lee, Michael McCahill, and Ma~Rosalyn Mesina.
\newblock Beyond the ‘{All} {Seeing} {Eye}’: {Filipino} {Migrant} {Domestic} {Workers}’ {Contestation} of {Care} and {Control} in {Hong} {Kong}.
\newblock {\em Ethnos}, 85(2):276--292, March 2020.

\bibitem{jonsen_using_2009}
Karsten Jonsen and Karen~A. Jehn.
\newblock Using triangulation to validate themes in qualitative studies.
\newblock 4(2):123--150, August 2009.

\bibitem{jupp_sage_2006}
Victor Jupp.
\newblock {\em The {SAGE} {Dictionary} of {Social} {Research} {Methods}}.
\newblock SAGE Publications, Ltd, 1 Oliver's Yard, 55 City Road, London England EC1Y 1SP United Kingdom, 2006.

\bibitem{keane_gdpr_2018}
Eddie Keane.
\newblock The {GDPR} and {Employee}'s {Privacy}: {Much} {Ado} but {Nothing} {New}.
\newblock {\em King's Law Journal}, 29(3):354--363, September 2018.

\bibitem{kelley_nutrition_2009}
Patrick~Gage Kelley, Joanna Bresee, Lorrie~Faith Cranor, and Robert~W. Reeder.
\newblock A "nutrition label" for privacy.
\newblock In {\em Proceedings of the 5th {Symposium} on {Usable} {Privacy} and {Security}}, pages 1--12, Mountain View California USA, July 2009. ACM.

\bibitem{knijnenburg_modern_2022}
Bart~P. Knijnenburg, Xinru Page, Pamela Wisniewski, Heather~Richter Lipford, Nicholas Proferes, and Jennifer Romano, editors.
\newblock {\em Modern {Socio}-{Technical} {Perspectives} on {Privacy}}.
\newblock Springer International Publishing, Cham, 2022.

\bibitem{koskei_role_2015}
B.~K. Koskei and C.~Simiyu.
\newblock Role of {Interviews}, {Observation}, {Pitfalls} and {Ethical} {Issues} in {Qualitative} {Research} {Methods}.
\newblock {\em Journal of Educational Policy and Entrepreneurial Research}, October 2015.

\bibitem{kraemer_informing_2019}
Martin~J. Kraemer, William Seymour, Reuben Binns, Max Van~Kleek, and Ivan Flechais.
\newblock Informing {The} {Future} of {Data} {Protection} in {Smart} {Homes}, June 2019.

\bibitem{lau_alexa_2018}
Josephine Lau, Benjamin Zimmerman, and Florian Schaub.
\newblock Alexa, {Are} {You} {Listening}?: {Privacy} {Perceptions}, {Concerns} and {Privacy}-seeking {Behaviors} with {Smart} {Speakers}.
\newblock {\em Proceedings of the ACM on Human-Computer Interaction}, 2(CSCW):1--31, November 2018.

\bibitem{lee_information_2016}
Linda Lee, JoongHwa Lee, Serge Egelman, and David Wagner.
\newblock Information {Disclosure} {Concerns} in {The} {Age} of {Wearable} {Computing}.
\newblock In {\em Proceedings 2016 {Workshop} on {Usable} {Security}}, San Diego, CA, 2016. Internet Society.

\bibitem{lee_electronic_2003}
Samantha Lee and Brian~H. Kleiner.
\newblock Electronic surveillance in the workplace.
\newblock {\em Management Research News}, 26(2/3/4):72--81, March 2003.

\bibitem{leenes_chapter_2018}
Ronald Leenes and Silvia~De Conca.
\newblock Chapter 10: {Artificial} intelligence and privacy—{AI} enters the house through the {Cloud}.
\newblock December 2018.

\bibitem{lipford_understanding_2008}
Heather~Richter Lipford, Andrew Besmer, Jason Watson, and Heather Lipford.
\newblock Understanding {Privacy} {Settings} in {Facebook} with an {Audience} {View}.
\newblock 2008.

\bibitem{liu_detecting_2018}
Tian Liu, Ziyu Liu, Jun Huang, Rui Tan, and Zhen Tan.
\newblock Detecting {Wireless} {Spy} {Cameras} {Via} {Stimulating} and {Probing}.
\newblock pages 243--255, Munich Germany, June 2018. ACM.

\bibitem{lupton_self-tracking_2014}
Deborah Lupton.
\newblock Self-tracking cultures: towards a sociology of personal informatics.
\newblock pages 77--86, Sydney New South Wales Australia, December 2014. ACM.

\bibitem{madden_opinion_2019}
Mary Madden.
\newblock Opinion {\textbar} {The} {Devastating} {Consequences} of {Being} {Poor} in the {Digital} {Age}.
\newblock {\em The New York Times}, April 2019.

\bibitem{malkin_what_2018}
Nathan Malkin, Julia Bernd, Maritza Johnson, and Serge Egelman.
\newblock "{What} {Can}'t {Data} {Be} {Used} {For}?": {Privacy} {Expectations} about {Smart} {TVs} in the {U}.{S}.
\newblock In {\em Proceedings 3rd {European} {Workshop} on {Usable} {Security}}, London, England, 2018. Internet Society.

\bibitem{malkin_privacy_2019}
Nathan Malkin, Joe Deatrick, Allen Tong, Primal Wijesekera, Serge Egelman, and David Wagner.
\newblock Privacy {Attitudes} of {Smart} {Speaker} {Users}.
\newblock {\em Proceedings on Privacy Enhancing Technologies}, 2019(4), October 2019.

\bibitem{manheim_artificial_2019}
Karl Manheim and Lyric Kaplan.
\newblock Artificial {Intelligence}: {Risks} to {Privacy} and {Democracy}.
\newblock 21, 2019.

\bibitem{gold_trending_2012}
Lev Manovich.
\newblock Trending: {The} {Promises} and the {Challenges} of {Big} {Social} {Data}.
\newblock In Matthew~K. Gold, editor, {\em Debates in the {Digital} {Humanities}}, pages 460--475. University of Minnesota Press, January 2012.

\bibitem{mare_smart_2020}
Shrirang Mare, Franziska Roesner, and Tadayoshi Kohno.
\newblock Smart {Devices} in {Airbnbs}: {Considering} {Privacy} and {Security} for both {Guests} and {Hosts}.
\newblock {\em Proceedings on Privacy Enhancing Technologies}, 2020(2):436--458, April 2020.

\bibitem{marky_you_2020}
Karola Marky, Sarah Prange, Florian Krell, Max Mühlhäuser, and Florian Alt.
\newblock “{You} just can’t know about everything”: {Privacy} {Perceptions} of {Smart} {Home} {Visitors}.
\newblock pages 83--95, Essen Germany, November 2020. ACM.

\bibitem{marky_i_2020}
Karola Marky, Alexandra Voit, Alina Stöver, Kai Kunze, Svenja Schröder, and Max Mühlhäuser.
\newblock ”{I} don’t know how to protect myself”: {Understanding} {Privacy} {Perceptions} {Resulting} from the {Presence} of {Bystanders} in {Smart} {Environments}.
\newblock In {\em Proceedings of the 11th {Nordic} {Conference} on {Human}-{Computer} {Interaction}: {Shaping} {Experiences}, {Shaping} {Society}}, pages 1--11, Tallinn Estonia, October 2020. ACM.

\bibitem{mathur_dark_2019}
Arunesh Mathur, Gunes Acar, Michael~J. Friedman, Eli Lucherini, Jonathan Mayer, Marshini Chetty, and Arvind Narayanan.
\newblock Dark {Patterns} at {Scale}: {Findings} from a {Crawl} of {11K} {Shopping} {Websites}.
\newblock {\em Proceedings of the ACM on Human-Computer Interaction}, 3(CSCW):1--32, November 2019.

\bibitem{mcdonough_bystanders_2019}
Oriana McDonough.
\newblock A {Bystander}'s {Dilemma}: {Participatory} {Design} {Study} of {Privacy} {Expectations} for {Smart} {Home} {Devices}.
\newblock 2019.

\bibitem{mchugh_interrater_2012}
Marry~L. McHugh.
\newblock Interrater reliability: the kappa statistic.
\newblock {\em Biochemia Medica}, pages 276--282, 2012.

\bibitem{naeini_privacy_2015}
Pardis~Emami Naeini.
\newblock Privacy {Expectations} and {Preferences} in an {IoT} {World}. {Open} {Access} {Media}. {USENIX}, June 2015.

\bibitem{nissenbaum_privacy_2009}
Helen Nissenbaum.
\newblock {\em Privacy in {Context}: {Technology}, {Policy}, and the {Integrity} of {Social} {Life}}.
\newblock Stanford University Press, January 2009.

\bibitem{nong_chain--thought_2024}
Yu~Nong, Mohammed Aldeen, Long Cheng, Hongxin Hu, Feng Chen, and Haipeng Cai.
\newblock Chain-of-{Thought} {Prompting} of {Large} {Language} {Models} for {Discovering} and {Fixing} {Software} {Vulnerabilities}, February 2024.

\bibitem{oladoyinbo_exploring_2024}
Tunbosun Oladoyinbo, Samuel Olabanji, Oluwaseun Olaniyi, Olubukola Adebiyi, Olalekan Okunleye, and Adegbenga Alao.
\newblock Exploring the {Challenges} of {Artificial} {Intelligence} in {Data} {Integrity} and its {Influence} on {Social} {Dynamics}.
\newblock {\em Asian Journal of Advanced Research and Reports}, 18:1--23, January 2024.

\bibitem{olaoye_enhancing_2024}
Favour Olaoye and Kaledio Potter.
\newblock Enhancing {Smart} {Home} {Automation} {Using} {AI}-{Driven} {IoT} {Systems}.
\newblock {\em Artificial Intelligence}, October 2024.

\bibitem{senior_blindspot_2009}
Shwetak~N. Patel, Jay~W. Summet, and Khai~N. Truong.
\newblock {BlindSpot}: {Creating} {Capture}-{Resistant} {Spaces}.
\newblock In Andrew Senior, editor, {\em Protecting {Privacy} in {Video} {Surveillance}}, pages 185--201. Springer London, London, 2009.

\bibitem{pearce_examining_2021}
Hammond Pearce, Benjamin Tan, Baleegh Ahmad, Ramesh Karri, and Brendan Dolan-Gavitt.
\newblock Examining {Zero}-{Shot} {Vulnerability} {Repair} with {Large} {Language} {Models}, December 2021.

\bibitem{pearman_user-friendly_2022}
Sarah Pearman, Ellie Young, and Lorrie~Faith Cranor.
\newblock User-friendly yet rarely read: {A} case study on the redesign of an online {HIPAA} authorization.
\newblock {\em Proceedings on Privacy Enhancing Technologies}, 2022(3):558--581, July 2022.

\bibitem{perera_big_2015}
Charith Perera, Rajiv Ranjan, Lizhe Wang, Samee~U. Khan, and Albert~Y. Zomaya.
\newblock Big {Data} {Privacy} in the {Internet} of {Things} {Era}.
\newblock {\em IT Professional}, 17(3):32--39, May 2015.

\bibitem{phiri_exponential_2023}
Mwalimu Phiri.
\newblock {EXPONENTIAL} {GROWTH} {OF} {DATA}, March 2023.

\bibitem{pierozzi_data_2018}
FILIPPO PIEROZZI.
\newblock Data {Power} {Europe} {GDPR} {Jurisdictional} {Reach} and the bid for an international unilateral standard, December 2018.

\bibitem{portnoff_somebodys_2015}
Rebecca~S. Portnoff, Linda~N. Lee, Serge Egelman, Pratyush Mishra, Derek Leung, and David Wagner.
\newblock Somebody's {Watching} {Me}?: {Assessing} the {Effectiveness} of {Webcam} {Indicator} {Lights}.
\newblock pages 1649--1658, Seoul Republic of Korea, April 2015. ACM.

\bibitem{ramya_pillars_2025}
R.~Ramya, S.~Priya, P.~Thamizhikkavi, and M.~Anand.
\newblock The {Pillars} of {AI} {Ethics}: {Transparency}, {Accountability}, and {Privacy}.
\newblock In {\em Responsible {Implementations} of {Generative} {AI} for {Multidisciplinary} {Use}}, pages 85--110. IGI Global, 2025.

\bibitem{ranade_cybert_2021}
Priyanka Ranade, Aritran Piplai, Anupam Joshi, and Tim Finin.
\newblock {CyBERT}: {Contextualized} {Embeddings} for the {Cybersecurity} {Domain}.
\newblock In {\em 2021 {IEEE} {International} {Conference} on {Big} {Data} ({Big} {Data})}, pages 3334--3342, December 2021.

\bibitem{shaelou_challenges_2023}
Stéphanie~Laulhé Shaelou and Yulia Razmetaeva.
\newblock Challenges to {Fundamental} {Human} {Rights} in the age of {Artificial} {Intelligence} {Systems}: shaping the digital legal order while upholding {Rule} of {Law} principles and {European} values.
\newblock {\em ERA Forum}, 24(4):567--587, December 2023.

\bibitem{shahid_data_2022}
Jahanzeb Shahid, Rizwan Ahmad, Adnan~K. Kiani, Tahir Ahmad, Saqib Saeed, and Abdullah~M. Almuhaideb.
\newblock Data {Protection} and {Privacy} of the {Internet} of {Healthcare} {Things} ({IoHTs}).
\newblock {\em Applied Sciences}, 12(4):1927, February 2022.

\bibitem{skyflow_how_2023}
Skyflow.
\newblock How to {Achieve} {Global} {Data} {Privacy} {Compliance} in 2023 - {Skyflow}, 2023.

\bibitem{stern_improving_2014}
Tziporah Stern and Nanda Kumar.
\newblock Improving privacy settings control in online social networks with a wheel interface.
\newblock {\em Journal of the Association for Information Science and Technology}, 65(3):524--538, 2014.

\bibitem{strauss_grounded_1997}
Anselm Strauss and Juliet~M. Corbin.
\newblock {\em Grounded {Theory} in {Practice}}.
\newblock SAGE, March 1997.

\bibitem{tabassum_i_2019}
Madiha Tabassum, Tomasz Kosinski, and Heather~Richter Lipford.
\newblock "{I} don't own the data": {End} {User} {Perceptions} of {Smart} {Home} {Device} {Data} {Practices} and {Risks}.
\newblock 2019.

\bibitem{tian_is_2023}
Haoye Tian, Weiqi Lu, Tsz~On Li, Xunzhu Tang, Shing-Chi Cheung, Jacques Klein, and Tegawendé~F. Bissyandé.
\newblock Is {ChatGPT} the {Ultimate} {Programming} {Assistant} -- {How} far is it?, April 2023.

\bibitem{tiwari_inbuilt_2023}
Mayank Tiwari, Manish Kumar, Arnab Srivastava, and Anu Bala.
\newblock Inbuilt {Chat} {GPT} {Feature} in {Smartwatches}.
\newblock In {\em 2023 {International} {Conference} on {Circuit} {Power} and {Computing} {Technologies} ({ICCPCT})}, pages 1806--1813, August 2023.

\bibitem{torre_framework_2016}
Ilaria Torre, Frosina Koceva, Odnan~Ref Sanchez, and Giovanni Adorni.
\newblock A framework for personal data protection in the {IoT}.
\newblock In {\em 2016 11th {International} {Conference} for {Internet} {Technology} and {Secured} {Transactions} ({ICITST})}, pages 384--391, December 2016.

\bibitem{trueman_structured_2015}
C~N Trueman.
\newblock Structured {Interviews} - {History} {Learning} {Site}. historylearningsite.co.uk. {The} {History} {Learning} {Site}., 2015.

\bibitem{tsai_turtleguard_2017}
Lynn Tsai, Primal Wijesekera, Joel Reardon, Irwin Reyes, Jung-Wei Chen, Nathan Good, Serge Egelman, and David Wagner.
\newblock {TurtleGuard}: {Helping} {Android} {Users} {Apply} {Contextual} {Privacy} {Preferences}.
\newblock 2017.

\bibitem{varadharajan_data_2016}
Vijayaraghavan Varadharajan and Shruti Bansal.
\newblock Data {Security} and {Privacy} in the {Internet} of {Things} ({IoT}) {Environment}.
\newblock {\em Connectivity Frameworks for Smart Devices}, pages 261--281, 2016.

\bibitem{vollmer_recital_2023}
Nicholas Vollmer.
\newblock Recital 32 {EU} {General} {Data} {Protection} {Regulation} ({EU}-{GDPR}), April 2023.

\bibitem{watkins_allen_workplace_2017}
Myria Watkins~Allen, Stephanie~J. Coopman, Joy~L. Hart, and Kasey~L. Walker.
\newblock Workplace {Surveillance} and {Managing} {Privacy} {Boundaries}.
\newblock 21(2):172--200, July 2017.

\bibitem{westin_opt_2019}
Fiona Westin and Sonia Chiasson.
\newblock Opt out of privacy or "go home": understanding reluctant privacy behaviours through the {FoMO}-centric design paradigm.
\newblock In {\em Proceedings of the {New} {Security} {Paradigms} {Workshop}}, pages 57--67, San Carlos Costa Rica, September 2019. ACM.

\bibitem{wilson_benefits_2017}
Wilson.
\newblock Benefits and risks of smart home technologies {\textbar} {Elsevier} {Enhanced} {Reader}, 2017.

\bibitem{wilson_smart_2015}
Charlie Wilson, Tom Hargreaves, and Richard Hauxwell-Baldwin.
\newblock Smart homes and their users: a systematic analysis and key challenges.
\newblock {\em Personal and Ubiquitous Computing}, 19(2):463--476, February 2015.

\bibitem{windl_saferhome_2022}
Maximiliane Windl, Alexander Hiesinger, Robin Welsch, Albrecht Schmidt, and Sebastian~S. Feger.
\newblock {SaferHome}: {Interactive} {Physical} and {Digital} {Smart} {Home} {Dashboards} for {Communicating} {Privacy} {Assessments} to {Owners} and {Bystanders}.
\newblock {\em Proceedings of the ACM on Human-Computer Interaction}, 6(ISS):680--699, November 2022.

\bibitem{woodruff_would_2014}
Allison Woodruff, Lauren Schmidt, Vasyl Pihur, Laura Brandimarte, Sunny Consolvo, and Alessandro Acquisti.
\newblock Would a privacy fundamentalist sell their {DNA} for \$1000... if nothing bad happened as a result? {The} {Westin} categories, behavioral intentions, and consequences.
\newblock 2014.

\bibitem{xue_who_2024}
Xiao Xue, Xinyang Li, Boyang Jia, Jiachen Du, and Xinyi Fu.
\newblock Who {Should} {Hold} {Control}? {Rethinking} {Empowerment} in {Home} {Automation} among {Cohabitants} through the {Lens} of {Co}-{Design}.
\newblock In {\em Proceedings of the {CHI} {Conference} on {Human} {Factors} in {Computing} {Systems}}, pages 1--19, Honolulu HI USA, May 2024. ACM.

\bibitem{yao_privacy_2019}
Yaxing Yao, Justin~Reed Basdeo, Oriana~Rosata Mcdonough, and Yang Wang.
\newblock Privacy {Perceptions} and {Designs} of {Bystanders} in {Smart} {Homes}.
\newblock {\em Proceedings of the ACM on Human-Computer Interaction}, 3(CSCW):1--24, May 2019.

\bibitem{zeng_end_2017}
Eric Zeng, Shrirang Mare, and Franziska Roesner.
\newblock End {User} {Security} \& {Privacy} {Concerns} with {Smart} {Homes}.
\newblock page~17, February 2017.

\bibitem{zeng_understanding_2017}
Eric Zeng and Franziska Roesner.
\newblock Understanding and {Improving} {Security} and {Privacy} in {Multi}-{User} {Smart} {Homes}: {A} {Design} {Exploration} and {In}-{Home} {User} {Study}.
\newblock page~19, 2017.

\bibitem{zheng_user_2018}
Serena Zheng, Noah Apthorpe, Marshini Chetty, and Nick Feamster.
\newblock User {Perceptions} of {Smart} {Home} {IoT} {Privacy}.
\newblock {\em Proceedings of the ACM on Human-Computer Interaction}, 2(CSCW):1--20, November 2018.

\bibitem{zhu_applying_2019}
Tianqing Zhu and Philip~S. Yu.
\newblock Applying {Differential} {Privacy} {Mechanism} in {Artificial} {Intelligence}.
\newblock In {\em 2019 {IEEE} 39th {International} {Conference} on {Distributed} {Computing} {Systems} ({ICDCS})}, pages 1601--1609, Dallas, TX, USA, July 2019. IEEE.

\end{thebibliography}

\begin{table*}[h!]

    \centering
    \caption{Status of Data Protection Regulation in World Leading Countries}
    \scalebox{1.0}
    {
\begin{tabular}{ccccc}
\hline
\textbf{Country} & \textbf{Data Law}                        & \textbf{Regulatory Body} & \textbf{Year} & \textbf{AI Regulation} \\ \hline
Turkey           & DPL\footref{turkeydpl}                                      & KVKK                        & 2016                  & No                    \\
Malaysia           & PDPA\footref{malasyialaw}                                      & Malaysia-PDP                        & 2011                  & No                    \\
Europe           & GDPR                                     & EDPS                        & 2018                  & No                    \\
USA              & {{} State   Dependent} & State Dependent          & 2022                  & No                                \\
China            & PIPL\footref{chinaPIPL}                                     & CAC                        & 2021                  & No                                \\
Brazil           & LGPD\footref{brazilLGPD}                                     & ANPD                      & 2021                  & No                    \\
India            & DPDP IT  Act\footref{indialaw}                                  & India DPA                     & 2023                  & No                                 \\
Canada           & PIPEDA Act\footref{canadaPIPEDA}                               & OPC                      & 2000                  & No                    \\ \hline
\end{tabular}
 }
    \label{tab:privacy-regworld}
\end{table*}

\begin{table*}[h!]
\centering
\caption{Summary of Categories and Themes}
\scalebox{0.8}{
\begin{tabular}{l|ll}
\hline
\multicolumn{1}{c}{\textbf{Category}}                   & \multicolumn{1}{c}{\textbf{Themes}}                         & \multicolumn{1}{c}{\textbf{Sub-Themes}} \\ \hline
                  & \multicolumn{1}{l|}{\textbf{Current ChatGPT Capabilities}} & Guidance and Explanation                \\ \cline{2-3} 
                                   & \multicolumn{1}{l|}{}                                       & Privacy Education                       \\ \cline{3-3} 
                                   & \multicolumn{1}{l|}{\textbf{AI for User Empowerment}}       & Power Dynamics and                      \\
\multirow{8}{*}{\textbf{Aspirations for AI-Enhanced Privacy}}                                   & \multicolumn{1}{l|}{\textbf{}}                              & Control Over Data                       \\ \cline{3-3} 
                                   & \multicolumn{1}{l|}{\textbf{}}                              & Real-Time Privacy Alerts                \\ \cline{2-3} 
   & \multicolumn{1}{l|}{\textbf{}}                              & Context-Aware                           \\
       & \multicolumn{1}{l|}{\textbf{AI for Personalization}}        & Privacy Recommendations                 \\ \cline{3-3} 
         & \multicolumn{1}{l|}{\textbf{and Adaptation}}                & Seamless Integration                    \\
                                   & \multicolumn{1}{l|}{\textbf{}}                              & with Smart Devices                      \\ \cline{2-3} 
\multicolumn{1}{l|}{}                                   & \multicolumn{1}{l|}{\textbf{AI for Ease of}}                & Simplified Privacy Controls             \\ \cline{3-3} 
\multicolumn{1}{l|}{}                                   & \multicolumn{1}{l|}{\textbf{Use and Accessibility}}         & Voice-Activated                         \\
\multicolumn{1}{l|}{}                                   & \multicolumn{1}{l|}{\textbf{}}                              & Privacy Management                      \\ \hline
\multicolumn{1}{l|}{}         & \multicolumn{1}{l|}{\textbf{Data Security}}                 & Encryption and Anonymization            \\ \cline{2-3} 
\multicolumn{1}{l|}{}                                   & \multicolumn{1}{l|}{\textbf{Compliance with}}               & Adherence to Laws                       \\ \cline{3-3} 
\multicolumn{1}{l|}{}                                   & \multicolumn{1}{l|}{\textbf{Privacy Regulations}}           & Data Residency  and Sovereign          \\ \cline{2-3} 
\multicolumn{1}{l|}{\textbf{AI Ethical, Security and}}     & \multicolumn{1}{l|}{\textbf{Transparency and}}              & Explainability of  AI Decisions         \\ \cline{3-3} 
\multicolumn{1}{l|}{\textbf{Regulatory Considerations}} & \multicolumn{1}{l|}{\textbf{Accountability}}                & Accountability in Case                  \\
\multicolumn{1}{l|}{}                  & \multicolumn{1}{l|}{\textbf{}}                              & of Data Breaches                        \\ \cline{2-3} 
\multicolumn{1}{l|}{}                                   & \multicolumn{1}{l|}{\textbf{Ethical Use of AI}}             & Minimizing AI Bias                      \\ \cline{3-3} 
\multicolumn{1}{l|}{}                                   & \multicolumn{1}{l|}{\textbf{}}                              & User Consent and Autonomy               \\ \hline
\end{tabular}
}
 \label{Table: Categories and Themes}
\end{table*}

\begin{table*}[h!]
 \caption{Demographic Information of Participants}
  \centering
  \scalebox{0.7}
  {
\begin{tabular}{lcccccccll}
\hline
\multicolumn{1}{c}{\textbf{P\#}} & \textbf{Gender} & \textbf{Country} & \textbf{Age} & \textbf{Degree} & \textbf{Occupation} & \textbf{Experience With} & \textbf{Experience With} & \textbf{Competence} & \textbf{Used Smart   Devices}   \\
\multicolumn{1}{c}{\textbf{}}    & \textbf{}       & \textbf{}            & \textbf{}          & \textbf{}       & \textbf{}           & \textbf{ChatGPT}         & \textbf{Smart Devices}   & \textbf{}           & \multicolumn{1}{c}{\textbf{}}   \\
\multicolumn{1}{c}{\textbf{}}    & \textbf{}       & \textbf{}            & \textbf{}          & \textbf{}       & \textbf{}           & \textbf{}         & \textbf{(Years)}         & \textbf{}           & \multicolumn{1}{c}{\textbf{}}   \\ \hline
PU01                             & Male            & UK                   & 20-29              & M.Sc.           & ICT                 & Yes                      & 4                        & Proficient          & Google Nest Audio               \\ \hline
PU02                             & Male            & UK                   & 30-39              & M.Sc.           & Telecom Engineer    & Yes                      & 7                        & Expert              & Google Home                     \\ \hline
PU03                             & Male            & UK                   & 40-49              & B.Sc.           & Nurse               & Yes                      & 9                        & Competent           &  Smart Thermostat           \\ \cline{1-2} \cline{4-10} 
PU04                             & Female          & UK                   & 30-39              & B.Sc.           & House Keeping       & Yes                      & 7                        & Competent           & Google Nest Audio               \\ \hline
PU05                             & Female          & UK                   & 30-39              & M.Sc.           & Baby Sitter         & Yes                      & 6                        & Competent           & Amazon Echo Dot                 \\ \hline
PU06                             & Female          & UK                   & 40-49              & B.Sc.           & Baby Sitter         & Yes                      & 8                        & Competent           & Amazon Echo Dot                 \\ \hline
AD01                             & Male            & UK                   & 20-29              & M.Sc.           & ICT                 & Yes                      & 3                        & Expert              & Amazon Echo Dot                 \\ \hline
AD02                             & Female          & UK                   & 20-29              & B.Sc.           & Teacher             & Yes                      & 4                        & Competent           & Google Home                     \\ \cline{2-10} 
AD03                             & Male            & UK                   & 30-39              & B.Sc.           & Doctor              & Yes                      & 6                        & Expert              & Amazon Echo Dot                 \\ \hline
AD04                             & Male            & UK                   & 40-49              & M.Sc.           & Accountant          & Yes                      & 7                        & Competent           & Amazon Echo Dot                 \\ \hline
AD05                             & Female          & UK                   & 20-29              & B.Sc.           & HR                  & Yes                      & 3                        & Competent           & Google Nest Audio               \\ \hline
AD06                             & Male            & UK                   & 20-29              & M.Sc.           & HR                  & Yes                      & 5                        & Competent           & Google Home                     \\ \hline
AI01                             & Male            & UK                   & 20-29              & B.Sc.           & AI Developer        & Yes                      & 4                        & Expert              & Smart Thermostat           \\ \hline
AI02                             & Male            & USA                  & 30-39              & M.Sc.           & AI Developer        & Yes                      & 6                        & Proficient          & Google Nest Audio               \\ \hline
AI03                             & Female          & Germany              & 40-49              & B.Sc.           & AI Developer        & Yes                      & 8                        & Expert              & Amazon Echo Dot                 \\ \hline
AI04                             & Female          & UK                   & 30-39              & PhD             & AI Developer        & Yes                      & 7                        & Expert              & Amazon Echo Dot                 \\ \hline
D01                              & Female          & UK                   & 30-39              & PhD             & UX Designer         & Yes                      & 8                        & Expert              & Google Nest Audio               \\ \hline
D02                              & Male            & USA                  & 40-49              & B.Sc.           & UX Designer         & Yes                      & 8                        & Expert              & Smart Thermostat           \\ \hline
D03                              & Female          & USA                  & 20-29              & M.Sc.           & UX Designer         & Yes                      & 4                        & Proficient          & Google Nest Audio               \\ \hline
R01                              & Male            & UK                   & 20-29              & B.Sc.           & Regulator           & Yes                      & 5                        & Competent           & Google Nest Audio               \\ \hline
R02                              & Female          & UK                   & 30-39              & PhD             & Regulator           & Yes                      & 7                        & Proficient          & Amazon Echo Dot                 \\ \hline
R03                              & Male            & France               & 40-49              & B.Sc.           & Regulator           & Yes                      & 9                        & Competent           & Google Nest Audio               \\ \hline
R04                              & Male            & Holland              & 30-39              & M.Sc.           & Regulator           & Yes                      & 6                        & Proficient          & Smart   Thermostat \\ \hline
\end{tabular}
}
\label{Table:Demographic Information of Participants}
\end{table*}

\end{document}